\pdfoutput=1
\documentclass[review, 3p]{elsarticle}

\usepackage{graphicx}
\usepackage{amsmath, amssymb, amsthm}
\usepackage{myconventions}
\usepackage{anslistings}
\usepackage{algorithm, algorithmic}
\usepackage{hyperref}
\usepackage{verbatim}
\usepackage{subcaption}
\usepackage{float}
\usepackage{graphicx}
\usepackage{units}
\usepackage{url}
\usepackage{enumitem}
\usepackage{multirow}
\usepackage{siunitx}

\newlength\myindent
\newcommand\bindent{%
  \begingroup
  \setlength{\itemindent}{\myindent}
  \addtolength{\algorithmicindent}{\myindent}
}
\newcommand\eindent{\endgroup}

\journal{Computers \& Mathematics with Applications}

\begin{document}

\begin{frontmatter}

\date{March 11, 2020}
\title{LE\textsc{o}P\textsc{art}: a particle library for FE\textsc{ni}CS}



%
\author[tudelft,crux]{Jakob M. Maljaars}
\ead{j.m.maljaars@tudelft.nl/maljaars@cruxbv.nl/jakobmaljaars@gmail.com}
\author[bpi_cam]{Chris N. Richardson \corref{cor1}}
\ead{chris@bpi.cam.ac.uk}
\author[carnegie]{Nathan Sime \corref{cor1}}
\ead{nsime@carnegiescience.edu}
\address[tudelft]{Environmental~Fluid~Mechanics, Faculty~of~Civil~Engineering~and~Geosciences, Delft~University~of~Technology, Stevinweg~1, 2600~GA Delft, The~Netherlands}
\address[crux]{CRUX~Engineering~BV, Pedro~de~Medinalaan~3, 1086~XK Amsterdam, The~Netherlands}
\address[bpi_cam]{University of Cambridge BP Institute, Madingley Road, Cambridge, CB3~0EZ, UK}
\address[carnegie]{Department of Terrestrial Magnetism, Carnegie Institution for Science, Washington, DC, USA}

\begin{abstract}
This paper introduces \leopart, an add-on for the open-source finite element software library \fenics\ to seamlessly integrate Lagrangian particle functionality with (Eulerian) mesh-based finite element (FE) approaches. \leopart~-~which is so much as to say: `Lagrangian-Eulerian on Particles'~-~contains tools for efficient, accurate and scalable advection of Lagrangian particles on simplicial meshes. In addition, \leopart\ comes with several projection operators for exchanging information between the scattered particles and the mesh and \textit{vice versa}. These projection operators are based on a variational framework, which allows extension to high-order accuracy. In particular, by implementing a dedicated PDE-constrained particle-mesh projection operator, \leopart\ provides all the tools for diffusion-free advection, while simultaneously achieving optimal convergence and ensuring conservation of the projected particle quantities on the underlying mesh.
A range of numerical examples that are prototypical to passive and active tracer methods highlight the properties and the parallel performance of the different tools in \leopart. Finally, future developments are identified. The source code for \leopart\ is actively maintained and available under an open-source license at
\url{https://bitbucket.org/jakob_maljaars/leopart}.
\end{abstract}


\end{frontmatter}


\section{Introduction}
\label{S:1}
Passive and active tracer methods find applications in a versatile range of engineering areas such as geophysical flows and environmental fluid mechanics \cite{Bagtzoglou1992, Delhez1999, Deleersnijder2001, Tackley2003}, experimental fluid mechanics \cite{Ouellette2006, Westerweel2013, Raffel2018}, and bio-medical applications \cite{Hathway2011, CohenStuart2011}, to name a few. 
In passive tracers methods, the Lagrangian particle motion is fully determined by the carrier flow and there is no feedback mechanism from the particles to the carrier flow. 
Such a feedback mechanism between the tracer particles and the carrier fluid is, however, typical to active tracer methods, in which particles mutually interact with the carrier fluid. 
To capture this interaction between the particles and the flow in numerical models, physical information which is carried by the tracer particles is used to estimate one or more additional source terms in the governing fluid equations. In a discrete setting, this typically requires the reconstruction of mesh fields from the scattered particle data when the fluid flow equations are solved using a mesh-based approach, such as finite difference (FD), finite volumes (FV), or finite elements (FE).
Application examples of active tracer methods include, among many others, the modelling of turbulent (reacting) flows \cite{Pope1985, Zhang2004}, and mantle convection problems \cite{Tackley2003, PvK1997, Christensen1994,Brandenburg2007}. 
Alternatively, the particle information can be used to solve the advective part of physical transport phenomena, resulting in so-called \textit{particle-mesh operator splitting schemes}, see, e.g., \cite{Edwards2012, Kelly2015}, and the earlier work by Maljaars et al. \cite{Maljaars2017, Maljaars2019, Maljaars2019_LNCSE} from which \leopart\ has evolved. Eliminating artificial dissipation by using Lagrangian particles for the discretization of the advection operator primarily motivates such methods, rendering the approach promising for simulating advection dominated flows \cite{Maljaars2017, Maljaars2019} or free-surface flows \cite{Kelly2015, Maljaars2017_JH}. 

In all the aforementioned methods and applications, particle-based and mesh-based discretization techniques essentially become intertwined. To render such a combination of Lagrangian particle methods in conjunction with mesh-based FD, FV, or FE solvers tractable for simulating practical engineering problems, a suite of dedicated, flexible and efficient tools is indispensable. 

The open-source library \leopart\ \cite{LEoPart} which is presented in this paper provides such a toolbox by integrating particle functionality in the open-source FE library \fenics\ \cite{Logg2012}. 
\leopart~-~which stands for `Lagrangian-Eulerian on Particles'~-~
contains utilities for efficiently and flexibly advecting and tracking a set of user-defined particles on simplicial meshes. In addition, \leopart\ contains a suite of tools for projecting scattered particle data onto the mesh and \textit{vice versa} in a high-order accurate, efficient and physically sound manner by implementing particle-mesh projection tools developed in the first author's recent work \cite{Maljaars2017, Maljaars2019}. It is particularly the latter feature which sets \leopart\ apart from the particle support in, e.g., PETSc \cite{petsc-user-ref} or the open-source particle library ASPECT \cite{aspect-doi-v2.1.0, Gassmoeller2018} that is built on top of the finite element package deal.II \cite{dealII91}.
%
%
The resulting combination of \leopart\ and \fenics\ is particularly suited for application to flow problems involving active or passive tracers, or to implement particle-mesh operator splitting schemes, as will be demonstrated by various numerical examples throughout. 

The paper is structured as follows. Section~\ref{sec:fenics_and_leopart} gives some background information on the encompassing \fenics\ library, and provides a helicopter view of \leopart. Section~\ref{sec:paricle_funcionality} describes the implementation of particles, as well as the advection and tracking of particles in \leopart. Section~\ref{sec:particle_mesh_interaction} details the available particle-mesh interaction strategies. Particular attention is paid to the PDE-constrained particle-mesh interaction in Section~\ref{sec:pde_constrained_pm}, which enables the reconstruction of conservative mesh fields from a set of moving particles. Section~\ref{sec:numerical_examples} illustrates some example applications, meanwhile paying attention to the performance and scaling properties of \leopart. Section~\ref{sec:conclusions} closes the paper by presenting conclusions and providing an outlook on future developments.
\section{Implementation in \fenics} \label{sec:fenics_and_leopart}
\subsection{A primer on FEniCS}
\label{S:2.1}

\fenics\ is a Finite Element (FE) framework, written in C++, with a Python interface. One of the major challenges of writing an FE code is that the computation which needs most user configuration is that of the local element tensor, which lies at the innermost part of the assembly loop. The local element tensor describes the matrix entries on each individual element, and relates to the physical equations of the system. FEniCS solves this problem by allowing the user to write these equations in a Domain Specific Language (DSL), which is then automatically compiled to C code to be called by the assembly loop.

In addition to simplifying the construction of the local element tensor, FEniCS makes it easier to run in parallel using MPI. Using the Python interface, running FEniCS code with MPI can be as simple as {\tt mpirun -n 32 python3 demo.py}. The mesh will be partitioned into 32 chunks, and the problem will be distributed across 32 processes for this job.

FEniCS consists of several components:
\begin{itemize}[noitemsep]
    \item{Unified Form Language (UFL) - providing the DSL component}
    \item{FEniCS Form Compiler (FFC) - which compiles the DSL into C code}
    \item{FIAT - the finite element tabulator}
    \item{DOLFIN - the main C++/Python package, which integrates I/O, assembly and solvers}
\end{itemize}
\noindent
The following Python script shows how the DOLFIN interface can simplify the expression of an FE problem and computation of its solution. Consider the weak form of the Poisson equation defined on the unit square $\Omega := (0, 1)^2$: find $u \in V$ subject to homogeneous Dirichlet boundary conditions, where $V$ is the appropriate solution space, such that
%
    \begin{equation}
        \areaIntegral{\Omega}{\nabla u\cdot\nabla v} = \areaIntegral{\Omega}{v \sin(x)}
        \quad \forall \; v  \in V.
    \end{equation}
%
\inputpython{./Snippets/poisson.py}
%
%

\noindent
Calling the {\tt solve()} method runs the form compiler (FFC) and compiles the symbolic expressions into C code, which is compiled and loaded into memory. A global matrix equation is then assembled using this generated code for the local element tensor, and finally an LU solver is called to solve the resulting system of equations. Whilst this example is simple and compact, many options exist to expand each part of the problem, for example by applying Dirichlet boundary conditions, or assembling matrix and vector separately, and choosing more sophisticated solvers. 

For larger problems, it is important to run in parallel using MPI. Mesh partitioning is performed using PT-SCOTCH or ParMETIS, and there is support for the HDF5 file format, which allows parallel access of large datasets. Third party libraries are used throughout, wherever possible: PETSc is the linear algebra backend of choice, with a large selection of parallel solvers available.

\noindent
FEniCS is an open-source package, and is available for various platforms. The latest information can be found on the project website \url{www.fenicsproject.org}.

\subsection{\leopart\ code structure}
\leopart\ is built on top of the \fenics\ package, and adds new concepts for the advection of particles on simplicial meshes, and the interaction between particles and the mesh. A central paradigm in the design of \leopart\ is that it serves as an add-on to \fenics, using the existing \fenics\ tool chain wherever possible. As a result all the \fenics\ functionality remains available for the user. 
In particular, \leopart\ is designed such that it seamlessly integrates with the mesh partitioning in \fenics\ facilitating parallelism using MPI. 

\begin{figure}[H]
    \centering
    \includegraphics[width=0.75\textwidth]{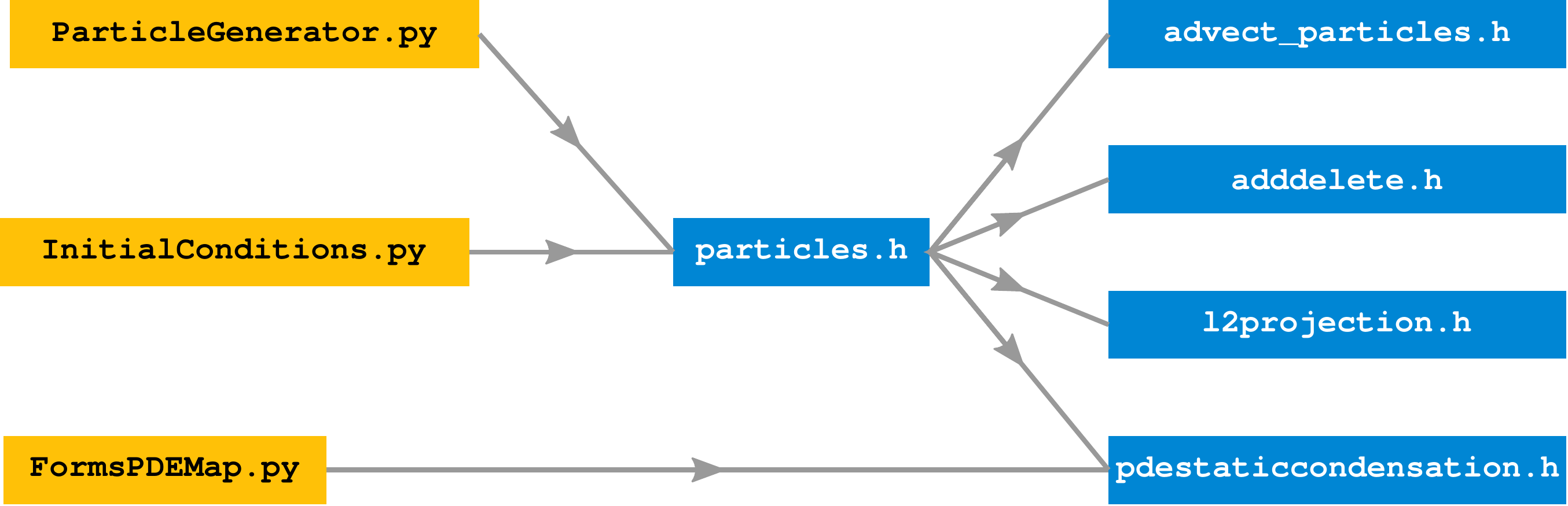}
    \caption{Code structure \leopart.}
    \label{fig:code_structure}
\end{figure}

To provide a fast and user friendly suite of tools, \leopart\ wraps C++ code in Python using \texttt{pybind}\footnote{\url{https://github.com/pybind/pybind11}} for the computationally demanding parts such as the particle advection and the matrix assembly. Particle pre-processing and post-processing is done in Python. Fig.~\ref{fig:code_structure} provides an overview of the different core components in \leopart. 

The remainder of this paper essentially discusses the different components from Fig.~\ref{fig:code_structure}. Particular attention is paid to the particles class, the advection of the particles, and the tools for the interaction between the particles and the mesh via $\ell^2$ projections or PDE-constrained projections.

As an aside, we note that \leopart\ also contains a Stokes solver, implementing the H(div) conforming hybridized discontinuous Galerkin (HDG) formulation from Rhebergen~\&~Wells \cite{Rhebergen2017, Rhebergen2018}. 
Implementation details of this solver are beyond the scope of this paper.

The source code for \leopart\ as well as the examples that are shown in what follows, are hosted under an open-source license at \url{https://bitbucket.org/jakob_maljaars/leopart}. 

\section{Particle Functionality} \label{sec:paricle_funcionality}
This section explains the implementation of particles in \leopart\ and discusses the particle advection and particle tracking strategy on simplicial meshes. 
\subsection{Particle initialization}
The \texttt{particles} class forms the backbone for dealing with the Lagrangian particles in \leopart. Operations such as particle advection and the particle-mesh projections require as input an instance of this \texttt{particles} class, see Fig.~\ref{fig:code_structure}. 

Each particle is assumed to have at least a spatial coordinate attached, which henceforth is  denoted by $\xp$ for a single particle. Moreover, it is presumed that particles always live in a spatial domain, denoted $\Omega$, so that the particle coordinate set is defined as 
%
	\begin{equation} \label{eq:coordinate_set}
	\mathcal{X}_t
	:= 
	\{\xp(t) \in \Omega\}_{p=1}^{N_p},
	\end{equation}
%
with $N_p$ the total number of particles. For notational convenience, we also make use of the index set of particles and the index set of particles hosted by cell $K$, at a fixed time instant $t$, which are defined as
%
	\begin{align} 
	\mathcal{S}_t &:= \defineSet{p\in\mathbb{N}}{\mathbf{x}_p(t) \in \mathcal{X}_t}, \label{eq:particle set definition} \\
	\mathcal{S}^{K}_t &:= \defineSet{p\in \mathbb{N}}{\mathbf{x}_p(t) \in \overline{K}, \hspace{3pt} \mathbf{x}_p(t) \in \mathcal{X}_t}. \label{eq:particle set definition_local}
	\end{align} 
%

Whereas carrying a spatial coordinate might be sufficient for passive particle tracing, additional properties, such as density, concentration, or momentum values, need to be attached to the particles for active particle tracing.
For a scalar and vector valued property, such particle quantities are defined as
%
    \begin{align} 
    	\Psi_t &:= \left\{ \psi_p(t) \in \mathbb{R} \right\}_{p=1}^{N_p}, \\
    	\mathcal{V}_t &:= \left\{ \mathbf{v}_p(t) \in \mathbb{R}^d  \right\}_{p=1}^{N_p},
    \end{align}
respectively, where $d=1,2,3$ is the spatial dimension.

%
\leopart\ provides a number of particle generators via \href{https://bitbucket.org/jakob_maljaars/leopart/src/master/source/ParticleGenerator.py}{\texttt{ParticleGenerator.py}} to generate a set of point coordinates. Most of the available particle generators create random point locations within a geometric object such as a rectangle (\texttt{RandomRectangle}), a circle (\texttt{RandomCircle}), and a sphere \texttt{RandomSphere}. These particle generators enable generating particles on parts of the meshed domain, but have the drawback that the point coordinates are generated on one processor and broadcasted to the other processors in parallel computations. 
This replication on MPI-ranks is circumvented in the \texttt{RandomCell} class, which generates random point coordinates within the simplicial (i.e. triangular or tetrahedral) cells of the mesh. For a tetrahedron, random barycentric coordinates within a cell are generated with the algorithm from \cite{Rocchini2000}, i.e.
\inputpython{./Snippets/random_bary.py}
Multiplying \texttt{s, t, u, v} by the vertex coordinates returns after summation a random point within a tetrahedral cell. Since \leopart\ inherits the domain decomposition of the mesh from DOLFIN, the \texttt{RandomCell} particle generator takes advantage of the mesh sub-domains, and only creates point coordinates that are within processor boundaries.

The set of Lagrangian particles which is formed by the coordinate set $
\mathcal{X}_t$, and an arbitrary number and ordering of scalar and/or vector valued particle quantities is used to instantiate the \texttt{particles} class in \leopart. 
Upon instantiation of this class, the hosting cell for a particle is found via a cell collision check that is available in \fenics\ via  \texttt{BoundingBoxTree::compute\_first\_entity\_collision}. As soon as the initial hosting cell is known, \leopart\ uses a more efficient algorithm for tracking moving particles on the mesh, see Section~\ref{sec:particle_tracking}. The coordinates of a point, together with the optional properties, constitute a \texttt{particle}, which is defined in \leopart\ as an array of \texttt{dolfin::Point}s, i.e.
\inputcpp{./Snippets/particle.cpp}
The first element in this array is always populated by the particle position. Using a \texttt{dolfin::Point} for the representation of the particle position allows to conveniently use other DOLFIN functionalities. However, this design choice also restricts the possibilities for the definition of the particle properties. A particle can, for example, neither carry vector valued properties for which the length exceeds three, nor can tensor properties be defined at the particles.

From the user's perspective, instantiating the \texttt{particles} class on a unit square mesh from a user-defined coordinate array, together with arrays for a scalar- and a vector-valued property is done as
\inputpython{./Snippets/particle_init.py}
For the sake of generality it is noted that the ordering and the length of the list with the particle quantities, i.e. \texttt{[}\texttt{psi\_p}, \texttt{v\_p}\texttt{]} in the above example is arbitrary.

\subsection{Particle advection} \label{sec:particle_advection}
Three different particle advection schemes are currently supported by \leopart. These advection schemes solve the system of ODEs: given a vector-valued velocity field $\mathbf{a}_h$, solve $\forall \;p \in \pset$
%
    \begin{subequations} \label{eq:ode_particl_advection}
    \begin{align}
        \frac{\text{d} \mathbf{x}_p}{\text{d} t} &= \mathbf{a}_h(\mathbf{x}_p, t) \\
        \frac{\text{d} \psi_p}{ \text{d} t} &= 0, \\
        \frac{\text{d}  \mathbf{v}_p}{ \text{d} t} &= 0,
    \end{align}
    \end{subequations}
%
where a particle can carry an arbitrary number of scalar- and/or vector-valued quantities that will stay constant throughout the particle advection. 
The \texttt{advect\_particles} class solves Eq.~\eqref{eq:ode_particl_advection} with a first-order accurate Euler forward method, and the two and three stage Runge-Kutta methods are available via the \texttt{advect\_rk2} and the \texttt{advect\_rk3} classes, respectively. The two multi-stage Runge-Kutta advection schemes inherit from the \texttt{advect\_particles} class, and a typical constructor for the latter reads
\inputcpp{./Snippets/advection_constructor.cpp}
This snippet shows that the particle advection classes require a \texttt{particles} instance, a velocity field specified in the \texttt{Function} and its corresponding \texttt{FunctionSpace}, and a \texttt{MeshFunction} for marking the boundaries, see Section~\ref{sec:particle_boundary_conditions}. A complete overview of all the particle advection constructors is found in \href{https://bitbucket.org/jakob_maljaars/leopart/src/master/source/cpp/advect_particles.h}{\texttt{advect\_particles.h}}.

\subsubsection{Cell-particle connectivity and particle relocation} \label{sec:cell_particle_connectivity}
Imperative for both the particle advection, as well as the particle-mesh interactions later on, is the evaluation of mesh fields at a potentially large number of points inside the domain. In order to do so, it must be known which cell is hosting the particle. At a meta-level, two options are available to fit this purpose. The first option is that each particle carries a reference to its hosting cell in the mesh. As soon as a particle crosses a cell boundary, this particle-to-cell reference is updated. Alternatively, a cell can be considered as a bucket filled with particles. A particle is removed from the cell's `particle bucket' as soon as it escapes the cell, and added to the receiving cell's particle bucket. 
Rather than a particle keeping track of its hosting cell, the bookkeeping is done at the cell level, i.e. each cell contains a list of particles.  

The latter method is used in \leopart, as this enables efficient evaluation of a mesh-field at the particle positions and allows to conveniently use the \fenics\ mesh partitioning for storing the particle data on the different processes. Central to the method is the cell-particle connectivity table, for which \leopart\ uses
\inputcpp{./Snippets/cell2part.cpp}
Related to the advection stage, that will be discussed in more detail below, this \texttt{\_cell2part} structure can be updated with the \texttt{particles::relocate} method, for which the declaration reads
\inputcpp{./Snippets/relocate.h}
This method is run once per advection step~-~or once per sub-step for the multistage advection schemes~-~and takes as input a relocation array \texttt{reloc} of particle indices (\texttt{pidx}) that should be relocated from one cell (\texttt{cidx}) to another (\texttt{cidx\_recv}). For each entry in the \texttt{reloc} array, \leopart\ either copies the particle to a particle collector if the particle escaped through a boundary or creates and pushes a particle to the receiving cell index via
\inputcpp{./Snippets/relocate_implementation.cpp}
Once this relocation is done, the particles that needed relocation are erased from the old cell index via \texttt{particles::delete\_particle}:
\inputcpp{./Snippets/del_particle.cpp}

In view of this relocation method, the following remarks are made:
\begin{itemize}[noitemsep]
    \item A potential improvement could be to let the cell-particle connectivity \texttt{\_cell2part} store a list of particle indices per cell rather than a list of \texttt{particle} objects, and store the particles within a process in a separate array. This obviates the need to create and erase a \texttt{particle} within \texttt{\_cell2part} when it escapes to another cell on the same process.
    \item The particle communication between processors via the collection step and a subsequent pushing step will be further detailed when discussing the implementation of internal boundaries in Section~\ref{sec:internal_boundaries}. 
    \item An efficient way of finding the new hosting cell \texttt{cidx\_recv} given the current host \texttt{cidx} is crucial from a performance perspective. The implementation of a fast particle tracking algorithm is discussed next.  
\end{itemize}

\subsubsection{Particle tracking}\label{sec:particle_tracking}
A challenge that is specifically related to the particle advection, is to efficiently keep track of the hosting cell for the Lagrangian particles in the unstructured simplicial mesh. Several procedures have been developed in literature, such as superposition of a coarse Cartesian mesh onto the unstructured mesh~\cite{Muradoglu2006}, the tetrahedral walk method~\cite{Lohner1990}, or methods based on barycentric interpolation~\cite{Maljaars2016}. An alternative method is the convex polyhedron method~\cite{Haworth2010}, which assumes that the mesh consists of convex polyhedral cells. This indeed is the case for the simplicial cells used in \fenics. For each \textit{facet} in the mesh, the midpoint, the unit normal, and the connectivity of the facet are pre-computed. Concerning the connectivity, a facet has two neighboring cells for facets internal to the mesh, or only one neighboring cell for exterior boundary facets and internal facets which are located on processor boundaries. From the perspective of a mesh cell, indicated by $\cell$, the sign of the facet unit normals is adapted so as to make sure that they are always outwardly directed.
\begin{figure}[H]
\centering
\includegraphics[width = 0.5\textwidth]{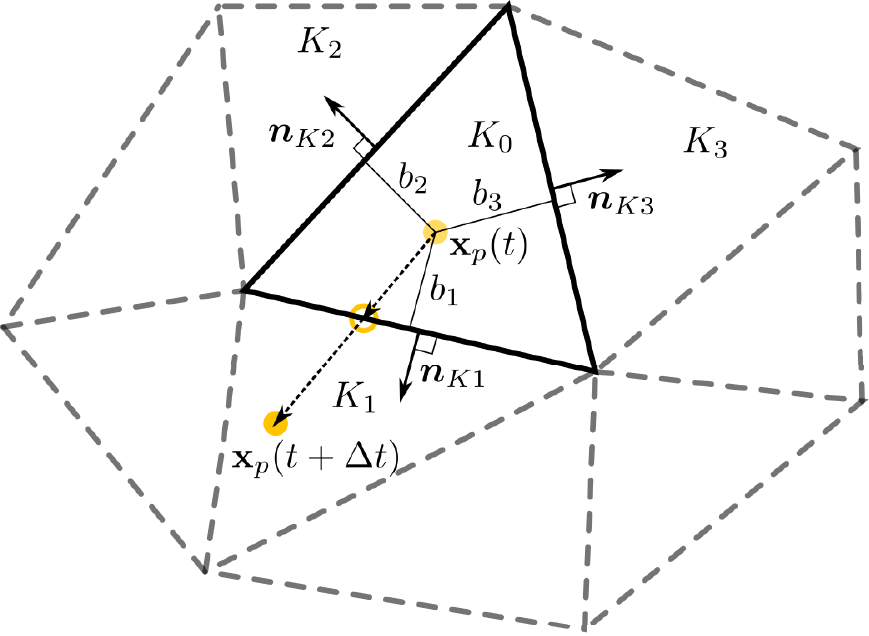}
\caption{Particle tracking using the convex polyhedron strategy.}
\label{fig:convex polyhedron}
\end{figure}

The convex polyhedron particle tracking then proceeds as follows for a particle $p$, which at time $t$ is located at $\mathbf{x}_p(t)$ within cell $K_0$, see Fig.~\ref{fig:convex polyhedron} for a principle sketch. Assume the particle has a velocity $\boldsymbol{a}\left(\mathbf{x}_p, t\right)$ and the time step used for advecting the particle is $\Delta t^{(0)} = \Delta t$, where the use of a superscript will become clear shortly. As the first step in the particle tracking algorithm, the time to intersect the i-th facet of element $K_0$ is computed as $\Delta t_i = b_i / ( \boldsymbol{a}\left(\mathbf{x}_p, t\right) \cdot \unitnormal_{K_i})$, with $ b_i $ the orthogonal distance between the particle and the facet with index $ i $. Next, the minimum, yet positive, time to intersection is computed as $\Delta t_{i_{\min}} = \min\{\max\left(0, \Delta t_i\right)\}$, with $i$ the indices of the neighboring cells. If $\Delta t_{i_{\min}} > \Delta t^{(k)}$ the particle is pushed to its new position using timestep $\Delta t^{(k)}$, and the time step is terminated by setting $\Delta t^{(k+1)} = 0$. If $\Delta t_{i_{\min}} < \Delta t^{(k)}$, the particle $p$ is pushed to the facet intersection $\mathbf{x}_p^{(k)}$ using $\Delta t_{i_{\min}}$, and the hosting cell is updated to facet $i_{\min}$, that is the facet with the index corresponding to $\Delta t_{i_{\min}}$. Furthermore, the time step is decremented to $\Delta t^{(k+1)} = \Delta t^{(k)} - \Delta t_{i_{\min}}$, after which the particle tracking continues until the time step for a particle has zero time remaining. Algorithm~\ref{algorithm:convex_polyhedron} contains the pseudo-code for the convex polyhedron particle tracking, using an Euler method for the particle advection.

The convex polyhedron procedure comes with a number of advantages, also pointed out in \cite{Haworth2010}. First of all, it is applicable to arbitrary polyhedral meshes, both in two and three spatial dimensions. Even though the current stable release of \fenics\ (2019.1.0) only supports simplicial meshes, this renders a generalization of the particle tracking scheme to other cell shapes straightforward, on the premise that cells are convex. Secondly, by marking the facets on the boundary of the domain, it is straightforward to detect if, when, and where a particle hits a specific boundary. 
This feature is useful when dealing with external boundaries, as well as internal boundaries, with the latter resulting from the mesh partitioning in parallel computations. Finally, the fraction of the time step spent in a certain cell is explicitly known in the convex polyhedron method. This can facilitate the updating of the particle velocity along its trajectory \cite{Popov2008}, although it is not further supported by the code in its present form. 
\begin{algorithm}[H]
\caption{Convex polyhedron particle tracking: pseudo-code for a single particle initially located in a cell $K$, using Euler forward for the particle advection.}
\label{algorithm:convex_polyhedron}
\small{
\begin{algorithmic}
\STATE $k \Leftarrow 0$
\STATE $\Delta t^{(k)} \Leftarrow \Delta t$
\STATE $\mathbf{x}_p^{(k)} \Leftarrow \mathbf{x}_p(t)$
\STATE Hosting cell: $\cell$
\WHILE{$\Delta t^{(k)} > 0$}
\STATE $k \Leftarrow k+1$
\STATE Time to facet intersection: $\Delta t_i = b_i / ( \boldsymbol{a}\left(\mathbf{x}_p^{0}, t\right) \cdot \boldsymbol{n}_{K_i})$, with $i$ the indices of the neighboring cells.
\STATE Minimum, yet positive time: $\Delta t_{i_{\min}} = \min\{\max\left(0, \Delta t_i\right)\}$, with $i$ the indices of the neighboring cells
\IF{ $\Delta t_{i_{\min}} > \Delta t^{(k)} $} 
\STATE Particle remains in cell $ \cell $
\STATE $\mathbf{x}_p^{(k)} \Leftarrow \mathbf{x}_p^{(k-1)} + \boldsymbol{a}\left(\mathbf{x}_p^{0}, t\right) \cdot \Delta t^{(k)}$
\STATE $\Delta t^{(k)}  \Leftarrow 0$
\ELSE 
\STATE Push particle to facet
\STATE $\mathbf{x}_p^{(k)} \Leftarrow \mathbf{x}_p^{(k-1)} + \boldsymbol{u}\left(\mathbf{x}_p^{0}, t\right) \cdot \Delta t_{i_{\min}}$
\STATE $\Delta t^{(k)} = \Delta t^{(k-1)} - \Delta t_{i_{\min}}$
\IF{ Facet has two neighboring cells}
\STATE Update hosting cell index: $K \Leftarrow K_{i_{\min}}$
\ELSE
\STATE Apply boundary condition, see Section~\ref{sec:particle_boundary_conditions} \label{line_algorithm}
\ENDIF 
\ENDIF 
\ENDWHILE
\STATE $\mathbf{x}_p(t + \Delta t) \leftarrow \mathbf{x}_p^{(k)}$
\end{algorithmic}
}
\end{algorithm}

\subsection{Boundary conditions at particle level} \label{sec:particle_boundary_conditions}
Apart from enforcing the boundary conditions on the background mesh, modifications at the particle level are also required when a particle hits a specific boundary. This event is detected when a particle is pushed to a facet having only one neighboring cell, see Line~\ref{line_algorithm} in Algorithm~\ref{algorithm:convex_polyhedron}. On the exterior boundary, the user can mark the different parts of the boundary as either being ``closed" (integer value 1), ``open" (integer value 2) or ``periodic" (integer value 3) via a \texttt{MeshFunction}, where this mesh function is passed to the particle advection class, see, for instance, the unit tests in \href{https://bitbucket.org/jakob_maljaars/leopart/src/master/unit_tests/test_2d_advect.py}{\texttt{test\_2d\_advect.py}}. Internal boundaries, i.e. facets that are on processor boundaries, are assigned an integer value of 0 by \leopart.

When a particle crosses either an internal, a periodic or an open boundary facet during the advection, we update the list of particle indices that needs to be relocated at the end of the time step as
\inputcpp{./Snippets/reloc_push_back.cpp}

Where \texttt{ci->index()} the cell index of the hosting cell, \texttt{i} the particle index within the hosting cell, and \texttt{numeric\_limits<unsigned int>::max()} the receiving cell index, with this value indicating that the particle cannot be tracked on the (partition of the) mesh. 

%
%
%
%
The implementation of the different particle boundary conditions is briefly discussed below. 
\subsubsection{Internal boundaries and particle communication} \label{sec:internal_boundaries}
At the end of each advection substep, \leopart\ tries to relocate the particles that escaped the old hosting cell by assigning them to the receiving cell via \texttt{particles::relocate}, see Section~\ref{sec:cell_particle_connectivity}. Particles that crossed a facet on an internal, an open or a periodic boundary , however, have a receiving cell index value of \texttt{numeric\_limits<unsigned int>::max()}. In this case, the particle is passed to the particle collector \texttt{particles::particle\_communicator\_collect} that prepares the particle for communication between the processors by I) finding the candidate host processor(s) via \texttt{BoundingBox::compute\_process\_collisions}, and II) appending the particle to the buffer that will be communicated to the candidate processor(s), i.e.
\begin{algorithm}[H]
\caption{Particle communication I:  \texttt{particles::particle\_communicator\_collect}.}
\label{algorithm:particle_communicator}
\small{
\begin{algorithmic}
\IF{\texttt{cidx\_recv} ==  \texttt{numeric\_limits<unsigned int>::max()}}
\STATE Find candidate hosting processor(s) \texttt{procs} via \texttt{dolfin::BoundingBox::compute\_process\_collisions}
\STATE Push a copy of the particle to the buffer \texttt{\_comm\_snd[\texttt{procs}]} that will be sent to the candidate processors
\ENDIF
\end{algorithmic}
}
\end{algorithm}
Once all the temporary copies of the particles that need communication are collected in \texttt{\_comm\_snd}, and after deleting these particles from the original hosting cell via \texttt{particles::delete\_particle}, the actual communication takes place via \texttt{particles::particle\_communicator\_push}. 
This communicates the particle copies to the candidate processor(s) via \texttt{MPI\_Alltoallv}, and checks whether the particle can be located on the candidate processor and in which cell via \texttt{BoundingBox::compute\_first\_entity\_collision}. If so, the communicated particle is recreated in the hosting cell on the candidate processor, if not, the other candidate processors are checked. 
In pseudocode, this particle relocation/communication strategy can be summarized as 
\begin{algorithm}[H]
\caption{Particle communication II:   \texttt{particles::particle\_communicator\_push}.}
\label{algorithm:particle_relocation_communications}
\small{
\begin{algorithmic}
\STATE Communicate \texttt{\_comm\_snd} (Algorithm~\ref{algorithm:particle_communicator}) to the buffer \texttt{comm\_rcv\_vec} on candidate processors via \texttt{MPI\_Alltoallv}
\FORALL{Particles \texttt{p} in \texttt{comm\_rcv\_vec}}
\IF{\texttt{p} collides with a cell in the candidate processor via \texttt{BoundingBox::compute\_first\_entity\_collision}}
\STATE Assign particle to receiving cell on candidate processor
\ENDIF
\ENDFOR
%
\end{algorithmic}
}
\end{algorithm}
Two remarks are made in view of this collection and pushing algorithm:
\begin{itemize}[noitemsep]
    \item The communication of particles between processors is as yet done via \texttt{MPI\_Alltoallv}. Although robust for large timesteps, this global communication might be somewhat inefficient since particles in general will move to processors that are close to the current processor. To exploit this, we will probably replace the communication by \texttt{MPI\_Neighbor\_alltoallv} in the future. This modification has, however, moderate priority since the numerical examples in Section~\ref{sec:numerical_examples} demonstrate that the particle advection usually represents a small fraction of run-time. 
    \item The receiving cell index for a particle that escaped through a facet on an open boundary is also set to \texttt{numeric\_limits<unsigned int>::max()} and hence sent to the particle collector \texttt{\_comm\_snd}. However, no new hosting cell is found in Algorithm~\ref{algorithm:particle_relocation_communications} for such a particle, and the particle is deleted as desired by the \texttt{particles::delete\_particle} that is executed in between the \texttt{particle\_communicator\_collect} and the \texttt{particle\_communicator\_push} method.
\end{itemize}

\subsubsection{Periodic boundaries}
When a particle crosses a facet which is marked as a periodic boundary, it should reappear at the opposing side of the domain. To implement this, facets on a periodic boundary, need to be matched against facets at the opposing side of the domain. This is taken care of in the advection class when the boundary is marked as a periodic boundary via a \texttt{MeshFunction}, and the coordinate-limits of the opposing boundaries are specified pairwise by the user. To illustrate this, a bi-periodic unit square domain is marked in \leopart, when using the forward Euler particle advection, as
\inputpython{./Snippets/mwe_periodic.py} 
Full code examples of how periodic boundaries are applied in 2D and 3D are found in \href{https://bitbucket.org/jakob_maljaars/leopart/src/master/tests/single_phase_ns/TaylorGreen_2D.py}{\texttt{TaylorGreen\_2D.py}} and \href{https://bitbucket.org/jakob_maljaars/leopart/src/master/tests/single_phase_ns/TaylorGreen_3D.py}{\texttt{TaylorGreen\_3D.py}}.

\subsubsection{Open boundaries and particle insertion/deletion}
At boundaries that are marked as open, particles either escape or enter the domain. When a particle escapes through an open boundary facet, it simply is deleted from the list of particles. Inflow boundaries, however, are less straightforward since new particles are to be created. This is done via the \texttt{AddDelete} class, which also allows a user to keep control over the number of particles per cell. 

\texttt{AddDelete} takes as arguments the particle class, a lower and an upper bound for the number of particles per cell, and a list of \fenics\ functions which are to be used for initializing the particle values. If a cell is marked as almost empty, i.e. the number of particles is lower than a preset lower bound for the number of particles per cell, the particle deficit is complemented by creating new particles. The locations for the new particles in their hosting cell are determined using a random number generator. 
To initialize the other particle quantities, two options are at the user's convenience: the particle value is either initialized based on a point interpolation from the underlying mesh field, or the particle value is assigned based on rounding-off the interpolated field value to a lower or upper boundary, i.e. 
\begin{equation}\label{eq:particle_init}
\psi_p
=
\begin{cases}
\psi_{\min} & \text{if} \qquad \xDiscreteScalar{\psi}(\mathbf{x}_p) \; \leq \; \frac{\psi_{\min} + \psi_{\max}}{2}, \\
\psi_{\max} & \text{if} \qquad \xDiscreteScalar{\psi}(\mathbf{x}_p) \; > \; \frac{\psi_{\min} + \psi_{\max}}{2}.
\end{cases}
\end{equation}
%
This feature is particularly useful when the particles carry binary fields, such as the density in two-fluid simulations.  

\begin{figure}[H]
\centering
\includegraphics[width=0.325\textwidth]{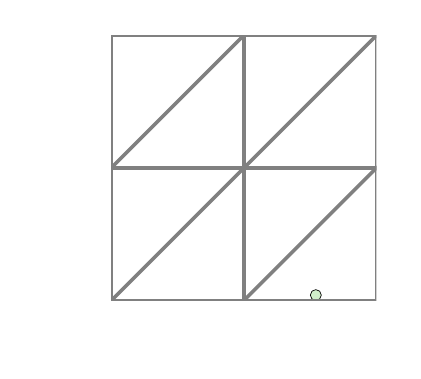}
\includegraphics[width=0.325\textwidth]{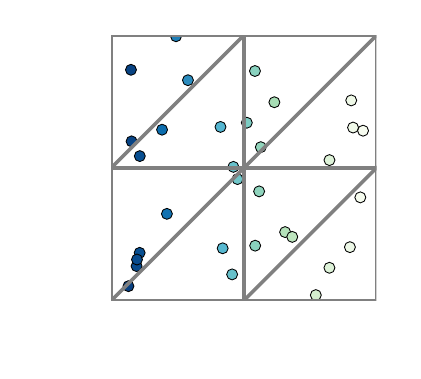}
\includegraphics[width=0.325\textwidth]{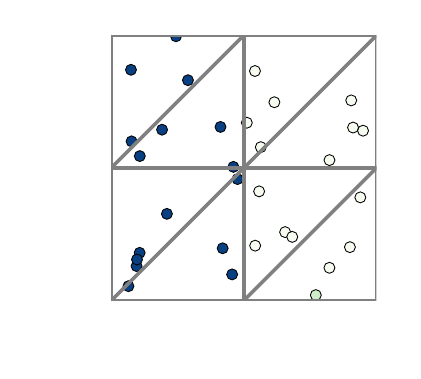}
\caption{Particle insertion: initial particle field (left), particle value assignment by interpolation (middle), binary particle value assignment (right).}
\label{fig:particle_insertion_deletion}
\end{figure}

The minimal example below demonstrates the \leopart\ implementation using the two options for particle insertion/deletion. Results are depicted in Fig.~\ref{fig:particle_insertion_deletion}.
\inputpython{./Snippets/mwe_insertion.py} 

The \texttt{AddDelete} class can also be used for keeping control over the maximum number of particles per cell by specifying the variable \texttt{p\_max} in the above presented code. If a cell in the \texttt{do\_sweep} method is marked to contain more particles than prescribed, the surplus of, say, $ m $ particles is removed by deleting $ m $ particles with the shortest distance to another particle in that cell. This procedure ensures that particles are removed evenly from the cell interior.

As a final remark: an upwind initialization of the particle value, i.e. initializing the particle value near open boundaries based on the value at the (inflow) boundary facet, is expected to be a useful feature not yet included in \leopart.
\subsubsection{Closed boundaries}
When a particle hits a closed boundary during the time step, the particle is reflected by setting the particle velocity to the reflected value, i.e.
%
\begin{equation}
	\mathbf{a}_h\left(\mathbf{x}_p, t\right) = \mathbf{a}_h\left(\mathbf{x}_p, t\right) - 2 \left(\mathbf{a}_h\left(\mathbf{x}_p, t\right)\cdot \unitnormal \right) \unitnormal,
\end{equation}
in which $\unitnormal$ the outwardly directed unit normal vector to a boundary facet. 

\section{Particle-mesh interaction} \label{sec:particle_mesh_interaction}
Obtaining mesh fields from the scattered particle data and updating the particle values from a known mesh field is essential to active tracer problems.
These particle-mesh interaction steps go by various names in the literature such as: `gather-scatter' steps \cite{Lohner1990, Chen2013}, `forward interpolation - backward estimation' \cite{Garg2007} or
`particle weighting' \cite{Jacobs2006}.

In line with our earlier work on particle-mesh schemes \cite{Maljaars2017, Maljaars2019, Maljaars2019_LNCSE}, the data transfer operators are consistently coined `particle-mesh projection' for the data
transfer from the set of scattered particles to the mesh, whereas the opposite route is indicated by `mesh-particle projection'. This convention reflects that the data transfer operators are perceived as projections between different spaces. More precisely, information needs to be projected from a \textit{particle space} onto a \textit{mesh space} and \textit{vice versa}. Adopting this point of view, it readily follows that the data transfer operations are auxiliary steps, which should not deteriorate accuracy, violate consistency, or compromise on conservation.   

To comply with these requirements \leopart\ adopts a variational approach to formulate the particle-mesh and the mesh-particle projections. An $\ell^2$ objective function is the starting point for deriving the mutual particle-mesh interactions. For a scalar-valued mesh field $\psi_h$ and a scalar-valued particle field $\psi_p$, this objective function reads
%
    \begin{equation} \label{eq: objective function}
        \min_{}	J:= \sum\limits_{p \in \mathcal{S}_t}^{} \frac{1}{2} \left( \psi_h(\mathbf{x}_p(t),t) - \psi_p(t) \right)^2,
    \end{equation}
%
where it remains to specify the minimizer, other than to say that either $\psi_h$ or $\psi_p$ are used to fit this purpose.  The implementation of the projection strategies which can be formulated based on Eq.~\eqref{eq: objective function} are further highlighted for a scalar quantity $\psi$ in the remainder of this section, and the projections for a vector-valued quantity follow the same path. More specifically, Section~\ref{sec:pm_projection} discusses the various particle-mesh projections available in \leopart, and in Section~\ref{sec:mp_projection} the available mesh-particle projections are discussed. 
The notation 
$\cellset := \{\cell\}$ is used throughout to indicate the set of disjoint cells $\cell$ that constitutes a meshing of the domain $\Omega$, and each cell $\cell$ has a boundary $\partial K$.

\subsection{Particle-mesh projections} \label{sec:pm_projection}
Common to the available particle-mesh projections in \leopart\ is the minimization of the objective function Eq.~\eqref{eq: objective function} with respect to an unknown mesh field $\psi_h$ given a known particle field $\psi_p$. For this, the function space in which $\psi_h$ is approximated must be defined.
To this end, \leopart\ conveniently exploits existing \fenics\ tools for defining arbitrary order polynomial function spaces. For reasons that become clear shortly, \leopart\ is tailored for projecting the particle data onto discontinuous function spaces at the mesh. In the scalar-valued setting these function spaces are defined by
%
    \begin{align}
    \fspacel   &:= \left\{w_h \in L^2(\cellset), \hspace{3pt} w_h\lvert_\cell \hspace{3pt}  \in P_k(\cell) \hspace{3pt} \forall \hspace{3pt} K \in \cellset \right\}, \label{eq:dg_wspace_local}
    \end{align}
%
where $\mathcal{T}$ is the partitioning of the domain $\Omega$ into a set of cells $\cell$, and $P_k(\cell)$ 
denotes the space spanned by Lagrange polynomials on $\cell$, where the subscript  $k\geq 0$ indicates the polynomial order.
\subsubsection{$\ell^2$-projection} \label{eq:l2_projection}
\label{sec:l2_pm}
With these definitions, the most elementary particle-mesh projection is found by minimizing Eq.~\eqref{eq: objective function} for $\psi_h \in \fspacel$, which results in the $\ell^2$-projection: given the particle values $\psi_p^{n} \in \pscalar$ and particle positions $\xp \in \mathcal{X}_t$, find $\xDiscreteScalar{\psi} \in \fspacel$ such that
%
\begin{align} 
	\sum\limits_{p \in \pset}^{} \left( \psi_h(\mathbf{x}_p(t),t) - \psi_p(t) \right)  w_h (\mathbf{x}_p(t)) &= 0 && \forall \hspace{3pt} w_h \in \fspacel. \label{eq:l2_proj_basic}
\end{align}

Given the definition for $\fspacel$ in Eq.~\eqref{eq:dg_wspace_local}, $\psi_h, w_h \in \fspacel$ are discontinuous across cell boundaries. Hence, Eq.~\eqref{eq:l2_proj_basic} is solved in a cellwise fashion, i.e.
%
    \begin{align} \label{eq:l2 proj cellsum}
    	\cellsum \sum\limits_{p \in \psetk }^{} \left( \psi_h(\mathbf{x}_p(t),t) - \psi_p(t) \right) w_h (\mathbf{x}_p(t)) &= 0 && \forall \hspace{3pt} w_h \in \fspacel,
    \end{align}
requiring the inversion of small, local matrices only, thus being amenable to an efficient, parallel implementation. 
%

The particle-mesh projection via the $\ell^2$ projection is implemented in \leopart\ in the \texttt{l2projection} class, which is instantiated as 
\inputcpp{./Snippets/l2projection_constructor.cpp}
in which the integer index \texttt{idx} indicates which particle property is projected. Projection onto a discontinuous space as in Eq.~\eqref{eq:l2 proj cellsum}, is done with the \texttt{project} method, which on the Python side can be invoked as
\inputpython{./Snippets/l2projection_python.py}
\leopart\ also allows projection of particle data onto a continuous Galerkin space~-~which leads to a global system for Eq.~\eqref{eq:l2_proj_basic}~-~by means of the \texttt{project\_cg} method
\inputpython{./Snippets/l2projection_cg_python.py}
\subsubsection{Bounded $\ell^2$-projection} \label{sec:l2_bounded_pm}
The minimization problem, Eq.~\eqref{eq:l2_proj_basic}, can be interpreted as a quadratic programming problem. This class of problems has been thoroughly analyzed in literature, and well-known techniques exist to extend these problems with equality, inequality, and box constraints, see e.g. \cite{Goldfarb1983} and references. In the context of the particle-mesh projection, imposing box constraints of the form 
\begin{equation}
l 
\leq \xDiscreteScalar{\psi} \leq 
u,
\end{equation}
can be particularly useful to ensure that the mesh field is bounded by $\left[l, u\right]$.

In \leopart, the box-constrained $\ell^2$ projection is implemented via a specialization of the \texttt{project} method, which can be invoked as
\inputpython{./Snippets/bounded_projection.py}
with \texttt{lb} and \texttt{ub} the user-specified lower- and upper bound, respectively.
At the backend, \leopart\ uses \texttt{QuadProg++}\footnote{\url{https://github.com/liuq/QuadProgpp}} for solving the box-constrained optimalization problem. The bounded $\ell^2$-projection is only available when projecting onto discontinuous function spaces. 

\subsubsection{PDE-constrained particle-mesh interaction} \label{sec:pde_constrained_pm}
The motivation for introducing Lagrangian particles~-~particularly when used as active tracers~-~is to conveniently accommodate advection. The particle-mesh projections presented in the preceding two sections, however, do not possess conservation properties. That is, initializing a particle quantity from an initial mesh field, advecting the particles, and subsequently reconstructing a mesh field from the updated particle positions with the (box-constrained) $\ell^2$-projection, results in a reconstructed mesh field with different integral properties. 
One way to conserve the mesh properties over the sequence of particle steps, is to keep track of the integral quantities on the mesh. 
This is accomplished by constraining the objective function for the particle-mesh projection, Eq.~\eqref{eq: objective function}, such that the reconstructed field $\psi_h$ satisfies a hyperbolic conservation law. The resulting PDE-constrained particle-mesh projection, developed in \cite{Maljaars2019}, possesses local (i.e. cellwise) and global conservation properties, and essentially involves solving the minimization problem:
given $\psi_p$, find $\psi_h \in \fspacel$ 	
%
\begin{subequations} \label{eq:constrained_optimization_abstract}
\begin{align} \label{eq:l2_objective}
\min_{\xDiscreteScalar{\psi} \in \fspacel} J = 
\sum_{p}^{} \frac{1}{2}\left( \psi_h(\mathbf{x}_p) - \psi_p\right)^2 
\end{align}
such that:
\begin{align}
&
\frac{\partial \xDiscreteScalar{\psi}}{\partial t}
+ \nabla \cdot \left( { \mathbf{a}\xDiscreteScalar{\psi} } \right)
= 0  \label{eq:constraint} \\
& +\text{BC's}
\end{align}
\end{subequations}
\hspace{2pt} is satisfied in a weak sense. For brevity, only periodic boundaries or boundaries 
with vanishing normal velocity (i.e. $\mathbf{a}\cdot \unitnormal = 0$) are considered in this paper. For a more elaborate discussion on other boundary conditions in Eq.~\eqref{eq:constrained_optimization_abstract}, reference is made to \cite{Maljaars2019}.

By casting the strong form of the constraint into a weak form by multiplying Eq.~\eqref{eq:constraint} with a Lagrange multiplier field $\xDiscreteScalar{\lambda} \in T_h$~-~with $T_h$ defined in \ref{app:pde_constrained_pm}, Eq.~\eqref{eq:lagrange_space}~-~and after applying integration by parts, the PDE-constrained optimization problem amounts to finding the stationary points of the Lagrangian functional
%
\begin{multline} \label{eq:scalar_lagrangian-functional}
	\mathcal{L}(\psi_h, \bar{\psi}_h, \lambda_h) 
	= 
	\sum_{p}^{} \frac{1}{2}\left( \psi_h(\mathbf{x}_p(t), t) - \psi_p(t)\right)^2 
	+ 
	\sum\limits_{K}^{}\lineIntegral{\partial K}{\frac{1}{2}\beta \left( \bar{\psi}_h - \psi_h \right)^2} 
	+
	\cellsum \areaIntegral{\cell}{\frac{1}{2} \zeta \lVert \nabla \psi_h \rVert^2}
	\\
	+ 
	\areaIntegral{\Omega}{\partialDt{\psi_h}\lambda_h } 
	- 
	\sum\limits_{\cell}^{} \areaIntegral{\cell}{ \mathbf{a} \psi_h \cdot \GradScalar{\lambda_h} } 
    + 
    \sum\limits_{K}^{} \lineIntegral{\partial K }{ \mathbf{a}\cdot \mathbf{n} \bar{\psi}_h \lambda_h} 
	,  
\end{multline}
%
in which the first term at the right-hand side is similar to the objective function in Eq.~\eqref{eq:l2_objective}. The second line in Eq.~\eqref{eq:scalar_lagrangian-functional} is a weak statement of the constraint equation, Eq.~\eqref{eq:constraint}. Furthermore, $\beta > 0$ is a small penalty parameter introduced to establish a coupling between $\psi_h$, and the control variable $\bar{\psi}_h$, where this control variable is defined on the facets of the cell via the trace space $\bar{W}_{h}$ from Eq.~\eqref{eq:wspace_global}, analogous to the flux variable in HDG methods, see, e.g., \cite{Labeur2012, Wells2011, Nguyen2009, Rhebergen2017}. Finally, $\zeta$ is a parameter which penalizes gradients, where this parameter is set to zero for smooth problems, and is only invoked when steep gradients in the mesh solution are to be expected, see Section~\ref{sec:slotted_disk}. 

A more in-depth interpretation of Eq.~\eqref{eq:scalar_lagrangian-functional} and analysis of the optimality system resulting after taking variations with respect to $\left(\psi_h, \lambda_h, \bar{\psi}_h \right) \in \left(W_h, T_h, \bar{W}_{h} \right)$, can be found in \cite{Maljaars2019}. \ref{app:pde_constrained_pm} provides a summary of the resulting variational forms in the fully-discrete setting, yielding a $3\times3$ block system at the element level, see Eq.~\eqref{eq:block_system_cell}.

\leopart\ implements the PDE-constrained particle-mesh projection via the \texttt{PDEStaticCondensation} class. The weak forms provided by \texttt{FormsPDEMap.py} are used to instantiate this class. Using notations similar to Eq.~\eqref{eq:block_system_cell}, a Python implementation may read
\inputpython{./Snippets/PDEStaticCondensation_python.py}
Assembly of the matrices and vectors is done via the \texttt{assemble} method
\inputpython{./Snippets/PDEStaticCondensation_assemble.py}
which serves a two-fold purpose: first of all it computes the element contributions for each cell $\cell$ in the $ 3 \times 3 $ block matrix
%
\begin{equation} \label{eq:block_system_cell}
	\begin{bmatrix}
	\boldsymbol{M}_p + \boldsymbol{N} 	& \boldsymbol{G}(\theta)	& \boldsymbol{L} \\
	\boldsymbol{G}(\theta)^\top  		& \boldsymbol{0}   			& \boldsymbol{H} \\
	\boldsymbol{L}^\top  				& \boldsymbol{H}^\top 		& \boldsymbol{B} 
	\end{bmatrix}
	\begin{bmatrix}
	\boldsymbol{\psi}^{n+1} \\ 
	\boldsymbol{\lambda}^{n+1} \\ 
	\boldsymbol{\bar{\psi}}^{n+1} \\
	\end{bmatrix}
	=
	\begin{bmatrix}
	\boldsymbol{\chi}_p \boldsymbol{\psi}_p^{n} \\ 
	\boldsymbol{G}(1-\theta)^\top \boldsymbol{\psi}^{n} \\
	\boldsymbol{0}\\
	\end{bmatrix},
\end{equation}
%
where the different contributions readily follow from \ref{app:pde_constrained_pm}, Eq.~\eqref{eq:discrete_optimality-adv-diff}.

Secondly, the \texttt{assemble} method assembles the local contributions into a global matrix-vector system. Since $\psi_h$ and $\lambda_h$ are local to a cell, the resulting global system of equations is expressed in terms of the flux variable $\bar{\psi}_h$ only. 
That is, the \texttt{assemble} method assembles the global system as follows
%
\begin{multline} \label{eq:global_system_pde}
\bigwedge_K
\left(
\mathbf{B}
-
\begin{bmatrix}
\mathbf{L} \\ 
\mathbf{H}
\end{bmatrix}^\top
\begin{bmatrix}
\mathbf{M}_p + \mathbf{N} & \mathbf{G} (\theta) \\
\mathbf{G}(\theta)^\top & \mathbf{0} 
\end{bmatrix}^{-1}
\begin{bmatrix}
\mathbf{L}\\
\mathbf{H}
\end{bmatrix}
\right)
\boldsymbol{\bar{\psi}}^{n+1}\\
=
-
\bigwedge_K
\begin{bmatrix}
\mathbf{L} \\ 
\mathbf{H}
\end{bmatrix}^\top
\begin{bmatrix}
\mathbf{M}_p + \mathbf{N} & \mathbf{G}(\theta) \\
\mathbf{G}^\top(\theta) & \mathbf{0} 
\end{bmatrix}^{-1}
\begin{bmatrix}
\boldsymbol{\chi}_p \boldsymbol{\psi}_p^{n} \\ 
\boldsymbol{G}(1-\theta)^\top \boldsymbol{\psi}^{*,n} \\
\end{bmatrix},
\end{multline}
%
in which the wedge $\bigwedge$ denotes assembly of the cell contributions into the global matrix, where this requires the inversion of a small saddle-point problem for each cell $\cell$ independently, so that the assembly procedure is amenable to a fast parallel implementation. 

The method \texttt{solve\_problem}
\inputpython{./Snippets/PDEStaticCondensation_solve.py}
solves the resulting global system, Eq.~\eqref{eq:global_system_pde}, for $\boldsymbol{\bar{\psi}}^{n+1}$. The \texttt{solver} and \texttt{preconditioner} can be specified by the user and defaults to the MUltifrontal Massively Parallel sparse direct Solver (MUMPS). In addition to solving the global problem, the \texttt{solve\_problem} method also applies the back substitution
%
\begin{equation} \label{eq:backsubstitution}
\begin{bmatrix}
\boldsymbol{\psi}^{n+1} \\ 
\boldsymbol{\lambda}^{n+1} \\ 
\end{bmatrix}
=
\begin{bmatrix}
\mathbf{M}_p + \mathbf{N} & \mathbf{G}(\theta) \\
\mathbf{G}^\top(\theta) & \mathbf{0} 
\end{bmatrix}^{-1}
\left(
\begin{bmatrix}
\boldsymbol{\chi}_p \boldsymbol{\psi}_p^{n+1} \\ 
\boldsymbol{G}(1-\theta)^\top \boldsymbol{\psi}^{*,n}
\end{bmatrix}
-
\begin{bmatrix}
\mathbf{L}\\
\mathbf{H}
\end{bmatrix}
\boldsymbol{\bar{\psi}}^{n+1}
\right),
\end{equation}
%
for obtaining the local unknowns $\boldsymbol{\psi}^{n+1}$ and (optionally) the Lagrange multiplier unknowns $\boldsymbol{\lambda}^{n+1}$. 

The sequence of steps for instantiating, assembling and solving the PDE-constrained particle mesh-projection with \leopart\ can be summarized in the algorithm:

\begin{algorithm}[H]
\caption{PDE-constrained projection algorithm}
\label{algorithm:pde_projection}
\small{
\begin{algorithmic}
\STATE Instantiate \texttt{PDEStaticCondensation}
\STATE Assemble with \texttt{PDEStaticCondensation::assemble}:
\bindent
\STATE Global matrix $\mathbf{A}_g$ = 0
\STATE Global vector $\mathbf{f}_g$ = 0
\FORALL{cells K in mesh}
    \STATE Assemble local contributions $\mathbf{N}, \mathbf{G}, \mathbf{L}, \mathbf{H}, \mathbf{B}$ from Eq.~\eqref{eq:block_system_cell} with \texttt{dolfin::LocalAssembler}
    \STATE Assemble particle contributions $\mathbf{M}_p, \boldsymbol{\chi}_p$ with 
    \texttt{particles::get\_particle\_contributions}
    \STATE Use \texttt{Eigen::inverse} to compute
    \begin{equation*}
        \begin{bmatrix}
            \mathbf{M}_p + \mathbf{N} & \mathbf{G}(\theta) \\
            \mathbf{G}^\top(\theta) & \mathbf{0} 
        \end{bmatrix}^{-1}
    \end{equation*}
    \STATE Add local contributions to global matrix: $\mathbf{A}_g \overset{+}{=} \text{LHS(Eq.~\eqref{eq:global_system_pde})}$
    \STATE Add local contribution to global vector $\mathbf{f}_g \overset{+}{=} \text{RHS(Eq.~\eqref{eq:global_system_pde})}$
\ENDFOR
\eindent
\STATE Solve using \texttt{PDEStaticCondensation::solve}:
\bindent
\STATE $\boldsymbol{\bar{\psi}}^{n+1} \Leftarrow \mathbf{A}_g^{-1} \mathbf{f}_g$
\FORALL{cells in mesh}
\STATE Compute $\boldsymbol{\psi}^{n+1}, \boldsymbol{\lambda}^{n+1}$ by backsubstitution in Eq.~\eqref{eq:backsubstitution}.
\ENDFOR
\eindent
%
\end{algorithmic}
}
\end{algorithm}

\subsection{Mesh-particle projection} \label{sec:mp_projection}
The mesh-particle projections, for updating particle properties from a given mesh field, also take the objective functional Eq.~\eqref{eq: objective function} as their starting point. Contrary to the particle-mesh projections, the particle-field is the unknown, so that the objective function needs minimization with respect to the particle field $\psi_p$, for the projection of a scalar-valued quantity.
Performing the minimization results in: given $\xDiscreteScalar{\psi} \in \fspacel $, find $\psi_p \in \Psi_t$ such that 
%
\begin{align}
	\sum\limits_{p\in \pset}^{} \left(\psi_h \left(\mathbf{x}_p(t),t\right) - \psi_p(t) \right) \; \delta \psi_p  = 0 && \forall \hspace{3pt} p \in \pset.
\end{align}
%
Since this equation must hold for arbitrary variations $\delta \psi$, the particularly simple result for the mesh-particle projection becomes
%
\begin{align} \label{eq:mesh-particle_semi-discrete}
	\psi_p(t) = \psi_h(\mathbf{x}_p(t),t)  && \forall \hspace{3pt} p \in \mathcal{S}_t,
\end{align}
%
i.e. particles values are obtained via interpolation of the mesh field. Interpolating a mesh field to particles is done in \leopart\ via the \texttt{interpolate} method in the \texttt{particle} class, i.e.
\inputpython{./Snippets/interpolate_python.py}
An interpolation overwrites the particle quantities with the interpolated mesh values. However, one of the assets of combining a high resolution particle field with a comparatively low-resolution mesh field is that the particle field may provide sub-grid information to the mesh \cite{Snider2001, McDermott2008, Popov2008}. In order to take advantage of this, the particles need to have a certain degree of independence from the mesh field. Analogous to the FLIP method \cite{Brackbill1986}, this is achieved by \textit{updating} the particle quantities by projecting the \textit{change} in the mesh field rather than \textit{overwriting} particle quantities. 
For a scalar valued quantity, this incremental update reads
%
\begin{align} \label{eq:mesh-particle semidiscrete}
    \dot{\psi}_p & = \dot{\psi}_h(\xp) && \forall \hspace{3pt} p \in \mathcal{S}_t,
\end{align}
in which $\dot{\psi}_h \in \fspacel$ the time derivative of the mesh field.

A fully-discrete implementation of Eq.~\eqref{eq:mesh-particle semidiscrete} is implemented in \leopart\ using the $\theta$ method for the time discretization:
%
\begin{align}\label{eq:discrete_particle_update}
	\psi_p^{n+1} = 
	\psi_p^n 
	+ 
	\Delta t \left( (1- \theta) \xtDiscreteScalar{\dot{\psi}}{n}\left(\mathbf{x}_p^n\right) 
	+ 
	\theta \xtDiscreteScalar{\dot{\psi}}{n+1}\left(\mathbf{x}_p^{n+1}\right) \right) 
	&& \forall \hspace{3pt} p \in \mathcal{S}_t,
\end{align}
%
where $\Delta t$ the time step, $0 \leq \theta \leq 1$, and $\xtDiscreteScalar{\dot{\psi}}{n} \in W_h$ is defined as $
    \dot{\psi}_h^n = (\psi_h^n - \psi_h^{n-1})/ \Delta t.
$
Eq.~\eqref{eq:discrete_particle_update} is available in \leopart\ via the \texttt{increment} method in the \texttt{particles} class, and can be used as
\inputpython{./Snippets/increment_python.py}
Two closing remarks are made in view of this incremental update:
\begin{itemize}[noitemsep]
    \item For step 1, $\theta=1$ since $\dot{\psi}_h^0$ is usually not defined.
    \item The increment from the old time level, i.e. $\dot{\psi}_h^n\left(\mathbf{x}_p^n\right)$ is stored at the particle level between consecutive time steps, for efficiency reasons. This requires an additional slot on the particles, i.e. \texttt{dpsi\_p\_dt}. The integer array in the \texttt{increment} call indicates which particle slots are used for the incremental update, i.e. \texttt{p.increment(psih\_new, psih\_old, [1, 2], theta, step)}.  
\end{itemize}

\section{Example applications} \label{sec:numerical_examples}
This section demonstrates the performance of \leopart\ in terms of accuracy, conservation and computational run time for a number of advection-dominated problems. On a per-time-step basis, the particle-mesh approach typically comprises the following sequence of steps:
\begin{enumerate}[noitemsep]
\item Lagrangian advection of the particles, as outlined in Section~\ref{sec:particle_advection}. 
\item particle-mesh projection to project the scattered particle data onto an Eulerian mesh field using either the $\ell^2$-projection, discussed in Section~\ref{sec:l2_pm}, the bound constrained $\ell^2$-projection from Section~\ref{sec:l2_bounded_pm}, or the PDE-constrained projection, Section~\ref{sec:pde_constrained_pm}. 
\item (optional) solve the physical problem~-~e.g. a diffusion or Stokes problem~-~at the mesh, using the reconstructed mesh field as a source term or as an intermediate solution to the discrete equations.
\item (optional) update the particles given the solution at the mesh, using the mesh-particle interaction tools from  Section~\ref{sec:mp_projection}.
\end{enumerate}
Step 1 and 2 are sufficient to solve an advection problem at the particles and to test the reconstruction of mesh fields from the moving particles. The sequence of steps 1-4 can be used for active tracer modelling or a particle-mesh operator splitting for, e.g., the advection-diffusion equation or the incompressible Navier-Stokes equations, see also Maljaars \cite{Maljaars2017, Maljaars2019}. 
For all the examples presented below, reference is made to the corresponding computer code in the \leopart\ repository on Bitbucket.

\subsection{Translation of a periodic pulse}
As a straightforward, yet illustrative example, the translation of the sinusoidal profile 
\begin{equation}
    \psi(\mathbf{x},0) = \sin{2\pi x} \sin{2 \pi y}
\end{equation}
on the bi-periodic unit square is considered, in analogy to LeVeque \cite{LeVeque1996}.
Owing to its simplicity, this test allows to assess the accuracy and the convergence properties of the $\ell^2$- and the PDE-constrained particle-mesh projection. Furthermore, it is used to illustrate the performance of the scheme by means of a strong-scaling study. Test results can be reproduced by running 
\href{https://bitbucket.org/jakob_maljaars/leopart/src/master/tests/scalar_advection/SineHump_convergence.py}{\texttt{SineHump\_convergence.py}}
for the convergence study, and \href{https://bitbucket.org/jakob_maljaars/leopart/src/master/tests/scalar_advection/SineHump_hires.py}{\texttt{SineHump\_hires.py}} for the scaling study.

\begin{figure}[H]
    \centering
    \hfill
    \includegraphics{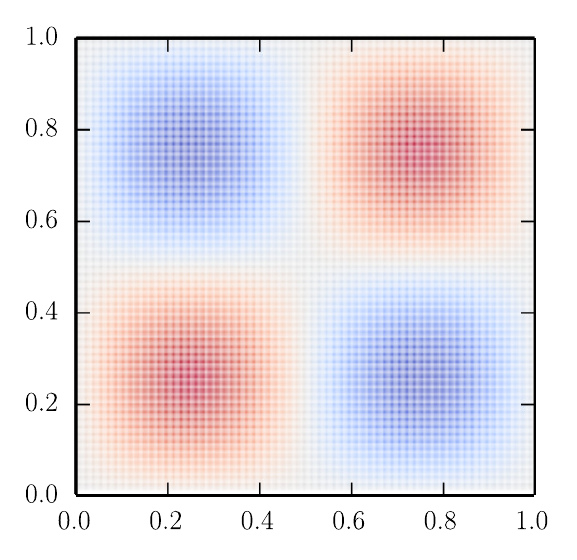}
    \hfill
    \includegraphics[height=6cm]{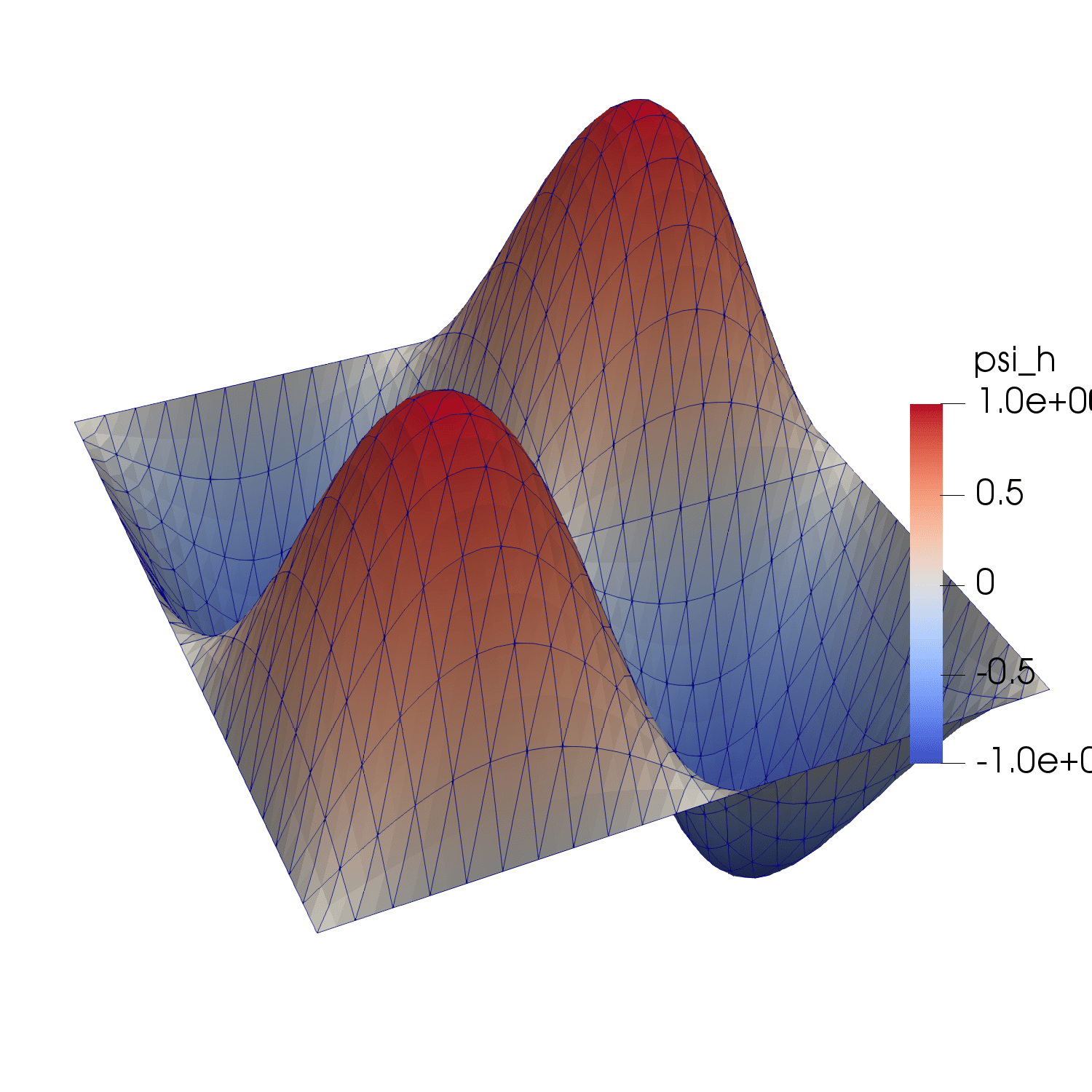}
    \hfill
    \caption{Sinusoidal Pulse: particle field (left) and the reconstructed solution at the mesh (right) using the PDE-constrained projection with polynomial order $k=2$ and a mesh containing 968 cells.}
    \label{fig:sine_hump_solution}
\end{figure}

In this example, the advective velocity field $\mathbf{a} = [1, 1]^\top$ is used, so that at $t = 1$ the initial data should be recovered. 

To investigate convergence, we consider a range of triangular meshes obtained by splitting a regular $n\times n$ Cartesian mesh into $2n^2$ triangles. We construct 5 different meshes with $n=(11,\, 22,\, 44,\, 88,\, 176)$, respectively.
%
%
Different polynomial orders $k=1, 2, 3$ are used for the discontinuous function space $W_h$, Eq.~\eqref{eq:dg_wspace_local}, onto which the particle data is projected. 
For the PDE-constrained projection, the polynomial order for the Lagrange multiplier space $T_h$, Eq.~\eqref{eq:lagrange_space}, is $l=0$ in all cases.
Particles are seeded in a regular lattice on the mesh, such that each cell contains approximately 15 particles, independent of the mesh resolution, see Fig.~\ref{fig:sine_hump_solution} as an example. An Euler scheme suffices for exact particle advection, and the time step corresponds to a CFL-number of approximately 1. Furthermore, in the PDE-constrained particle-mesh projection, the $\beta$-parameter is set to 1e-6, and $\zeta$ is set to 0. All computations use a direct sparse solver (\texttt{SuperLU}) to solve the global system of equations. Also, note that for this advection problem, the scalar valued property $\psi_p$, attached to each particle $p$, needs no updating and stays constant throughout the computation.

\begin{table}[H]
\centering
\caption{Translating pulse: overview of model runs with the associated $L^2$-error  $\lVert \psi - \psi_h \rVert$, and convergence rate at time $t=1$ for different polynomial orders $k$. For the PDE-constrained projection, the polynomial order in the Lagrange multiplier space is $l=0$ in all cases.} 
\label{tab:sine_profile_periodic}
\small{
\begin{tabular}{c r r r|c c|c c|c c}
\hline
\hline
& & & & \multicolumn{2}{c|}{$k = 1 $} & \multicolumn{2}{c|}{$k = 2 $} & \multicolumn{2}{c}{$k = 3 $} \\
Projection & $\Delta t $ & Cells & Parts.  &  Error & Rate & Error & Rate &  Error & Rate   \\
\hline
\multirow{ 5}{*}{$\ell^2$}
& 1e-1    &   242     &  3984     &   3.3e-2  &   - & 1.7e-3    &   - &   9.4e-5  &   - \\
& 5e-2    &   968     &  14542    &   8.3e-3  &   2.0 & 2.1e-4    &   3.0 &   5.9e-6  &   4.0 \\
& 2.5e-2  &   3872    &  57663    &   2.1e-3  &   2.0 & 2.7e-5    &   3.0 &   3.7e-7  &   4.0 \\
& 1.25e-2 &   15488   &  230428   &   5.2e-4  &   2.0 & 3.3e-6    &   3.0 &   2.3e-8  &   4.0 \\
& 6.25e-2 &   61952   &  921837   &   1.3e-4  &   2.0 & 4.1e-7    &   3.0 &   1.4e-9  &   4.0 \\
\hline
\multirow{ 5}{*}{PDE}
& 1e-1    &   242     &  3984     &   3.3e-2  &   - & 1.7e-3    &   - &   9.4e-5  &   - \\
& 5e-2    &   968     &  14542    &   8.3e-3  &   2.0 & 2.1e-4    &   3.0 &   5.9e-6  &   4.0 \\
& 2.5e-2  &   3872    &  57663    &   2.1e-3  &   2.0 & 2.7e-5    &   3.0 &   3.7e-7  &   4.0 \\
& 1.25e-2 &   15488   &  230428   &   5.2e-4  &   2.0 & 3.3e-6    &   3.0 &   2.3e-8  &   4.0 \\
& 6.25e-2 &   61952   &  921837   &   1.3e-4  &   2.0 & 4.1e-7    &   3.0 &   1.4e-9  &   4.0 \\
\hline
\hline 
\end{tabular} 
}
\end{table}

The accuracy of the method is assessed at $t=1$, using both the $\ell^2$-particle-mesh projection from Section~\ref{sec:l2_pm}, and the PDE-constrained projection from Section~\ref{sec:pde_constrained_pm} upon refining the mesh and the time step, see Table~\ref{tab:sine_profile_periodic}. Optimal convergence rates of order $k + 1$ are observed for both projections strategies, thus highlighting the accuracy of the particle-mesh projections in conjunction with particle advection. 

\begin{figure}[H]
\centering
    \begin{subfigure}{0.49\textwidth}
    	\centering
        \includegraphics{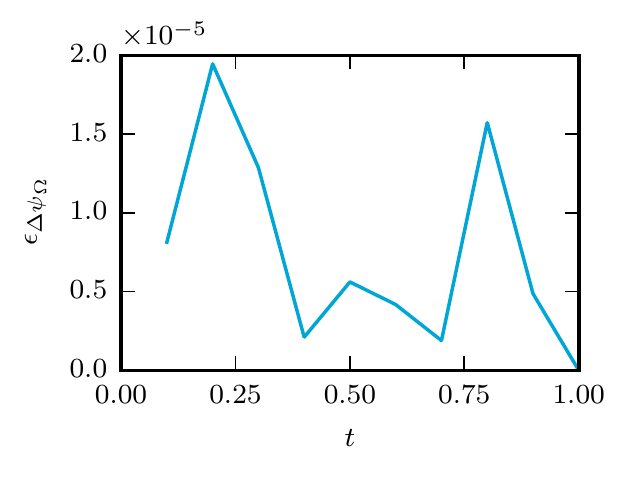}
        \caption{$\ell^2$-projection}
        \label{fig:sine_hump_l2}
    \end{subfigure}
    \begin{subfigure}{0.49\textwidth}
        \centering
        \includegraphics{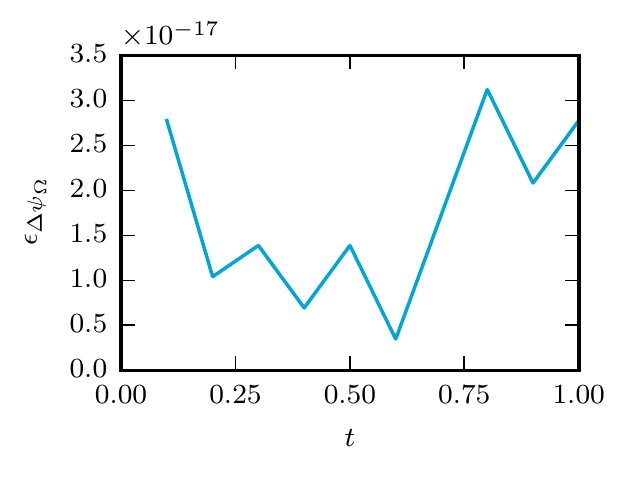}
        \caption{PDE-constrained projection}
        \label{fig:sine_hump_pde}
    \end{subfigure}
    \caption{Translating pulse: mass error over time for different particle-mesh projections.}
    \label{fig:sine_hump_mass_error}
\end{figure}

As reported in  Table~\ref{tab:sine_profile_periodic} the error levels for the $\ell^2$- and the PDE-constrained particle-mesh are similar. The difference between these projections becomes clear, however, by investigating the mass error
%
\begin{equation}\label{eq:area_error}
    \epsilon_{\Delta \psi_\Omega } =  \lvert \areaIntegral{\Omega}{ \left( \psi_h( \mathbf{x}, T) - \psi_h(\mathbf{x}, 0) \right)} \rvert,
\end{equation}
%
which is plotted as a function of time for the $\ell^2$-projection in Fig.~\ref{fig:sine_hump_l2}, and for the PDE-constrained projection in Fig.~\ref{fig:sine_hump_pde}. Evident from these figures is that the $\ell^2$-projection yields a nonzero mass error, whereas for the PDE-constrained projection the mass error is zero to machine precision.
\begin{figure}[H]
\centering
    \begin{subfigure}{0.49\textwidth}
    	\centering
        \includegraphics{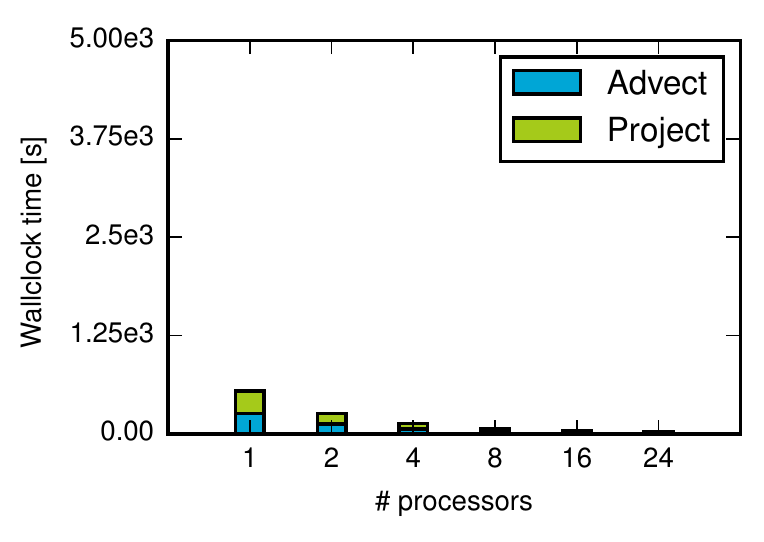}
        \caption{$\ell^2$-projection}
        \label{fig:sine_hump_l2_wallclock}
    \end{subfigure}
    \begin{subfigure}{0.49\textwidth}
        \centering
        \includegraphics{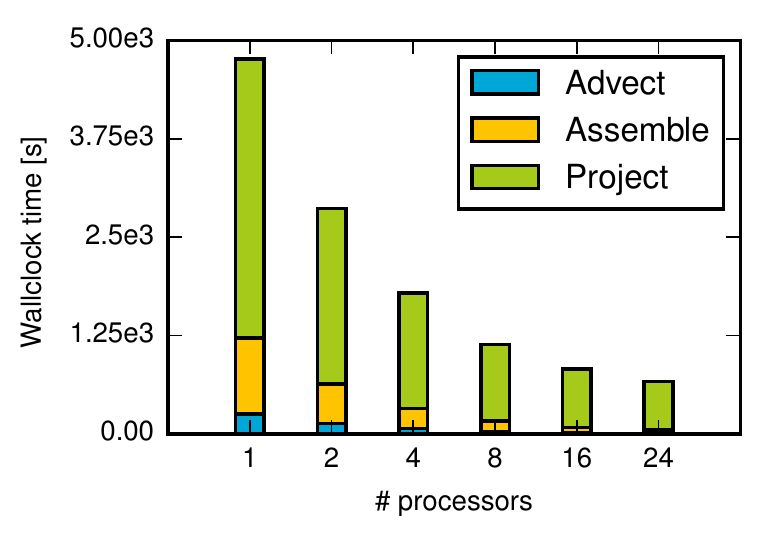}
        \caption{PDE-constrained projection}
        \label{fig:sine_hump_pde_wallclock}
    \end{subfigure}
    \caption{Translating sinusoidal hump: strong scaling study.}
    \label{fig:sine_hump_wallclock}
\end{figure}

The trade-off between the non-conservative $\ell^2$-projection and the conservative PDE-constrained is elucidated by investigating the computational times. Wallclock times for the high-resolution case~-~polynomial order $k=3$ with $61,952$ cells, $921,837$ particles and 160 time steps~-~run on different number of Intel Xeon CPU E5-2690 v4 processors are presented in Fig.~\ref{fig:sine_hump_wallclock}. 
Solving the global system for the PDE-constrained projection using a direct solver is computationally much more demanding compared to the (local) $\ell^2$-projection. This illustrates the need for an efficient iterative solver for the PDE-constrained projection step.

Table~\ref{tab:speed-up rotation sine hump} further investigates the scaling of the different components by summarizing the speed-up for the different tests relative to the run on one processor. The particle advection and the assembly step~-~with the latter only relevant for the PDE-constrained projection~-~exhibit excellent scaling properties, which is explained by the locality of these operations, i.e. these steps are performed cellwise. This also holds true for the $\ell^2$-projection, which amounts to a cellwise projection of the particle properties onto a discontinuous function space,  see Section~\ref{eq:l2_projection}. Clearly, the direct sparse solver for the PDE-constrained projection does not possess optimal scaling properties, which thus appears the limiting factor for the scalability of this step.
\begin{table}[H]
\centering
\caption{Translating sinusoidal hump: speed-up of the different model parts in parallel computations benchmarked against 1 processor run.}
\label{tab:speed-up rotation sine hump}
\begin{tabular}{c | c c c c c c | c c c c c c}
\hline
\hline
& \multicolumn{6}{c}{$\ell^2$-projection}  & \multicolumn{6}{|c}{PDE-constrained projection}  \\
\# Processors				& 1 	&	2	& 	4 	&	8	& 16 	& 24    & 1 	&	2	& 	4 	&	8	& 16 	& 24	\\
\hline
Advect particles	    & 1     & 2.1 &   4.1   & 8.0   & 15.1 & 22.4	& 1		& 1.9   & 	3.7 &   7.7 & 14.8	& 21.9	\\
Assembling 	            & -		& -   & 	-   &  -    & -	&  -		& 1     & 1.9   &   3.8 &   7.4 & 15.4  & 23.2  \\
Solve projection		& 1     & 2.1 &  4.3    & 8.2   & 16.1 & 25.1	& 1		& 1.6	& 	2.4	& 	3.6	& 4.8	&  5.8	 \\	
\hline
Total					&   1   &   2.1 &  4.2 &  8.1 &  15.6  & 23.8	& 1.    & 1.7   & 2.7  & 4.2    & 5.8   & 7.2	\\
\hline
\hline
\end{tabular}
\end{table}

\subsection{Slotted disk} \label{sec:slotted_disk}
Combining particle-based and mesh-based techniques appears particularly promising for applications in which sharp flow features are to be preserved. The solid body rotation of a slotted disk after Zalesak \cite{Zalesak1979} is a prototypical example of such problems, and often serves as a benchmark for interface tracking schemes, see \cite{Enright2002, Scardovelli2003}, among many others. We now use this test to demonstrate the various tools that \leopart\ offers for tracking sharp discontinuities in material properties, such as a density jump in immiscible multi-fluid flows.

The problem set-up is as follows. A disk with radius 0.2~-~from which a slot with a width of 0.1 and depth 0.2 is cut out~-~is initially centered at $(x,y)=(-0.15, 0)$ on the domain of interest $\Omega := \{(x,y)\hspace{1pt}\vert\hspace{1pt} x^2+y^2 \leq 0.5  \}$. This domain is triangulated into 14,464 cells on which 438,495 particles are seeded. The advective velocity field is given by
\begin{equation} \label{eq:solid_body_rotation}
    \mathbf{a} = \pi \left(-y, x\right)^\top.
\end{equation}
The time step is set to $0.02$, so that one full rotation is performed in 100 steps. The three-stage Runge-Kutta scheme, available via  the \texttt{advect\_rk3} class, is used for the paticle advection.
\begin{table}[h]
\centering
\caption{Slotted disk: overview of test cases, with $k$ the polynomial order in the function space $W_h$ (Eq~\eqref{eq:wspace_local}) and $l$ the polynomial order for the Lagrange multiplier space $T_h$ (Eq.~\eqref{eq:lagrange_space}) in the PDE-constrained projection.}
\label{tab:slotted_disk_overview}
\small{
\begin{tabular}{c|c c c c}
\hline
\hline 
\rule{0pt}{2.5ex}   
	& Particle-mesh projection	&	$k$	&	$l$	& $\zeta$	\\
\hline 
Case 1 & Bounded $\ell^2$	&	1	&	-	&	-   \\
Case 2 & PDE-constrained 	& 	1	& 	0	&  0   	\\
Case 3 & PDE-constrained    & 	1	& 	0	&  30   \\
\hline 
\hline 
\end{tabular}
}
\end{table}

Different particle-mesh projection strategies that are available in \leopart\ are investigated in this example, see Table~\ref{tab:slotted_disk_overview}. Note that Case~2 and Case~3 only differ in terms of the $\zeta$ parameter, see Eq.~\eqref{eq:scalar_lagrangian-functional}.
The computer code in \href{https://bitbucket.org/jakob_maljaars/leopart/src/master/tests/scalar_advection/SlottedDisk_rotation_l2.py}{\texttt{SlottedDisk\_rotation\_l2.py}} reproduces Case~1, the computer code for reproducing Case~2 and Case~3 is found in 
\href{https://bitbucket.org/jakob_maljaars/leopart/src/master/tests/scalar_advection/SlottedDisk_rotation_PDE.py}{\texttt{SlottedDisk\_rotation\_PDE.py}}. 
The test is run using 8 Intel Xeon CPU E5-2690 v4 processors.

\begin{figure}[h]
    \centering
    \begin{subfigure}{0.32\textwidth}
        \includegraphics{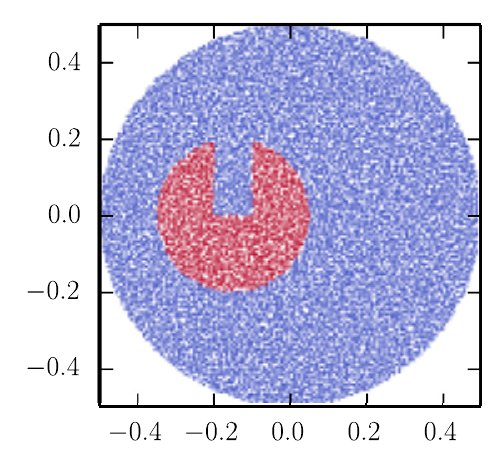} 
        \caption{$t=0$}
    \end{subfigure}
    \begin{subfigure}{0.32\textwidth}
        \includegraphics{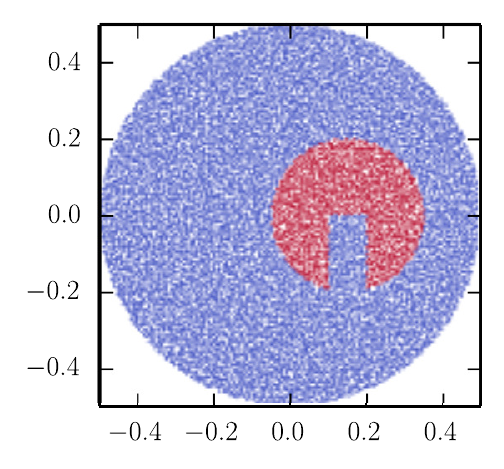}
        \caption{$t=1$}
    \end{subfigure}
    \begin{subfigure}{0.32\textwidth}
        \includegraphics{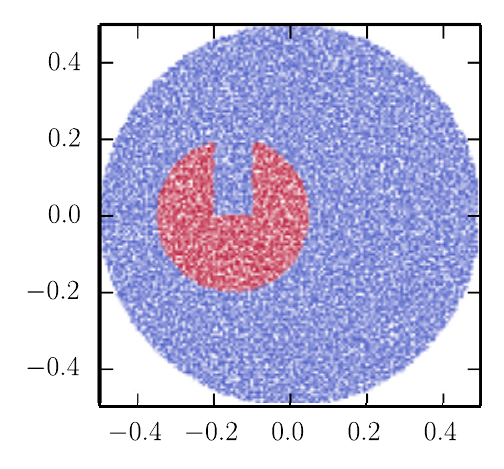} 
        \caption{$t=2$}
    \end{subfigure}
    \caption{Slotted disk: particle field (case independent) at different time instants. Plot shows every 25th particle for clarity, and color values range between 0 (blue) and 1 (red).}
    \label{fig:slotted_disk_particle_field}
\end{figure}

As for the previous example, there is no updating of the scalar valued property attached to a particle from the mesh-solution for this advection problem. Hence, sharp features at the particle level pertain, and the particle advection part for all the different cases is equal. Fig.~\ref{fig:slotted_disk_particle_field} shows the particle field at $t=0$ (initial field), $t=1$ (half rotation) and after a full rotation at $t=2$.  

\begin{figure}[H]
    \centering
    \begin{subfigure}{0.49\textwidth}
        \centering
        \includegraphics[width=5cm]{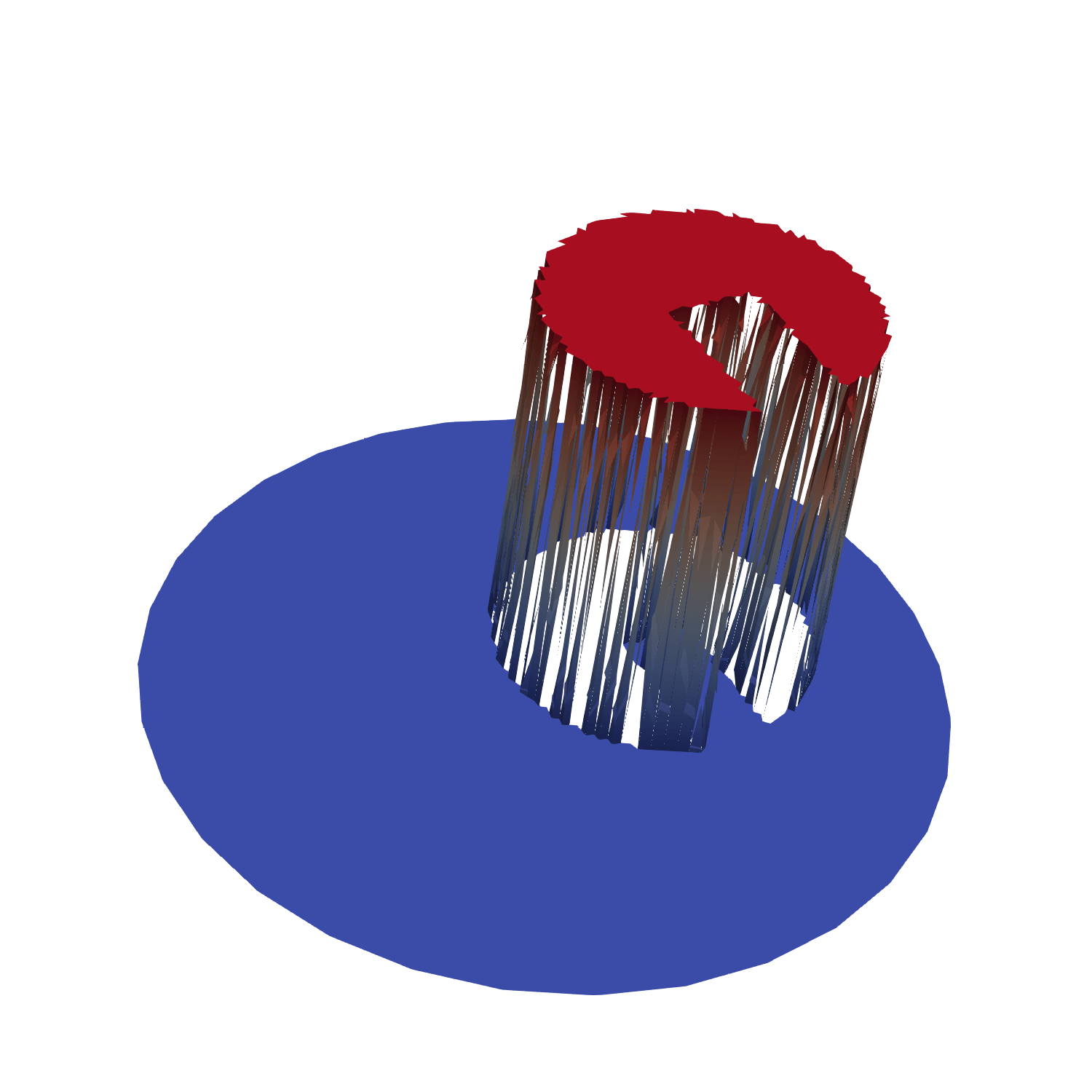}
        \caption{Case~1, $t=1$.}
    \end{subfigure}
    \begin{subfigure}{0.49\textwidth}
        \centering
        \includegraphics[width=5cm]{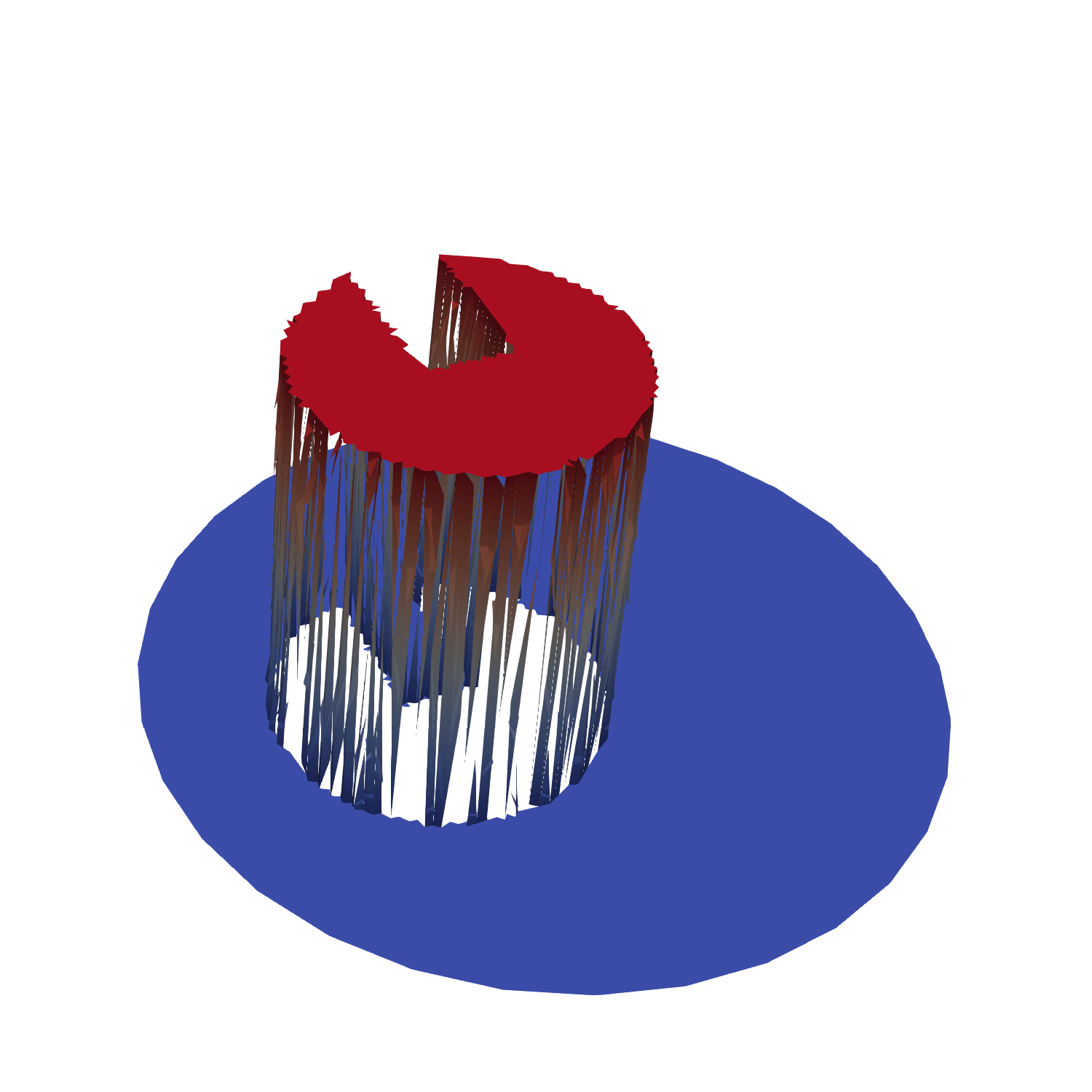}
        \caption{Case~1, $t=2$.}
    \end{subfigure}
    \begin{subfigure}{0.49\textwidth}
        \centering
        \includegraphics[width=5cm]{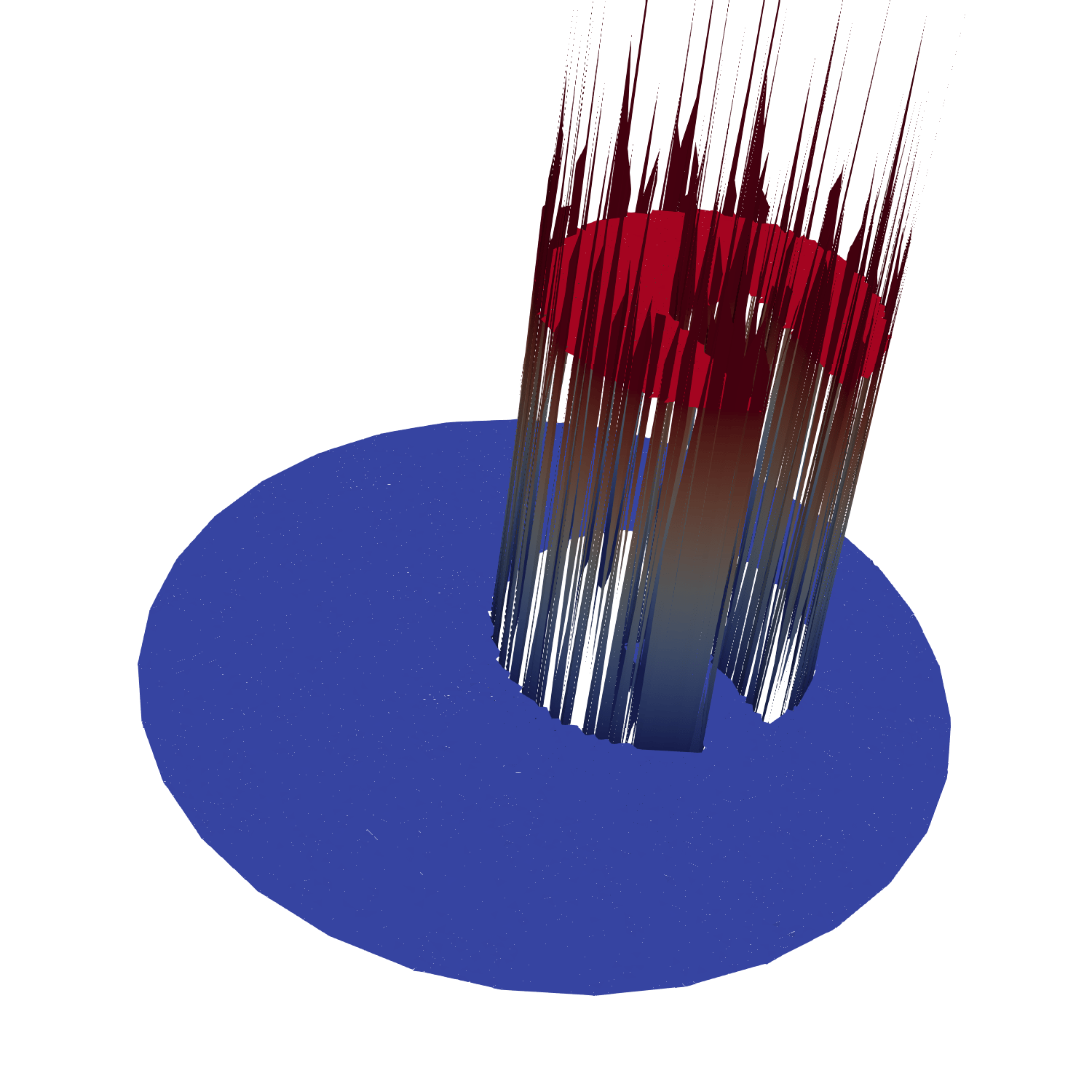}
        \caption{Case~2, $t=1$.}
    \end{subfigure}
    \begin{subfigure}{0.49\textwidth}
        \centering
        \includegraphics[width=5cm]{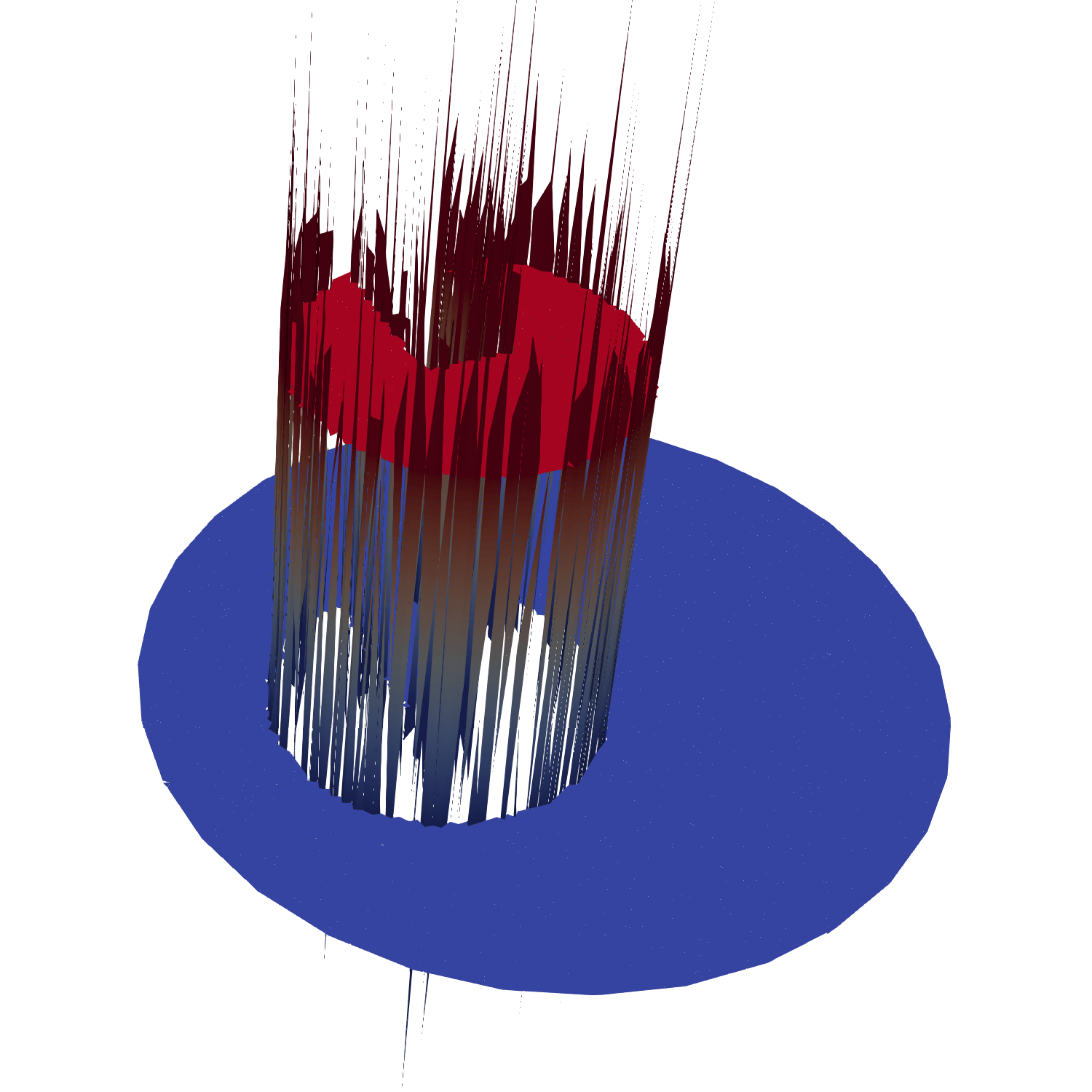}
        \caption{Case~2, $t=2$.}
    \end{subfigure}
    \begin{subfigure}{0.49\textwidth}
        \centering
        \includegraphics[width=5cm]{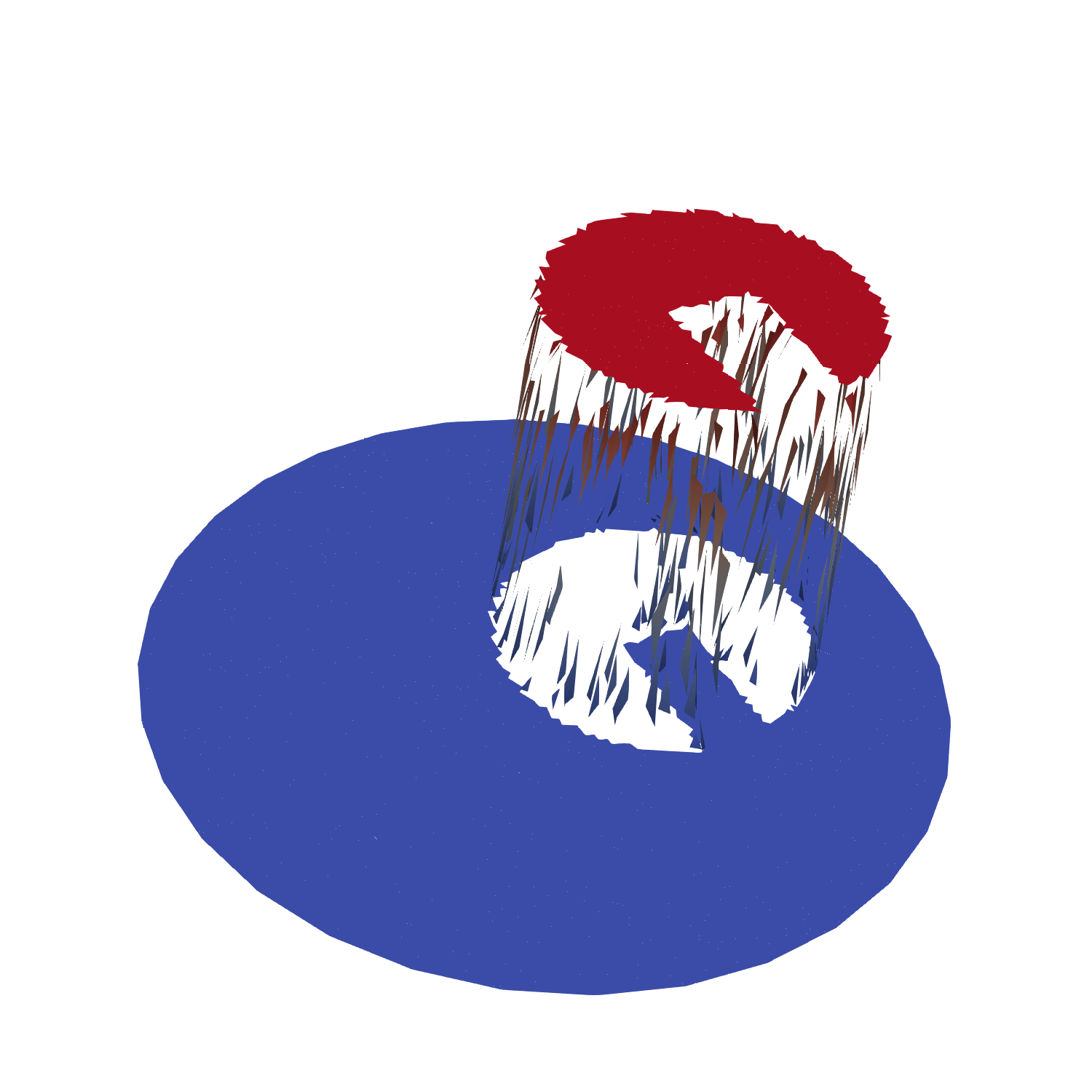}
        \caption{Case~3, $t=1$.}
    \end{subfigure}
    \begin{subfigure}{0.49\textwidth}
        \centering
        \includegraphics[width=5cm]{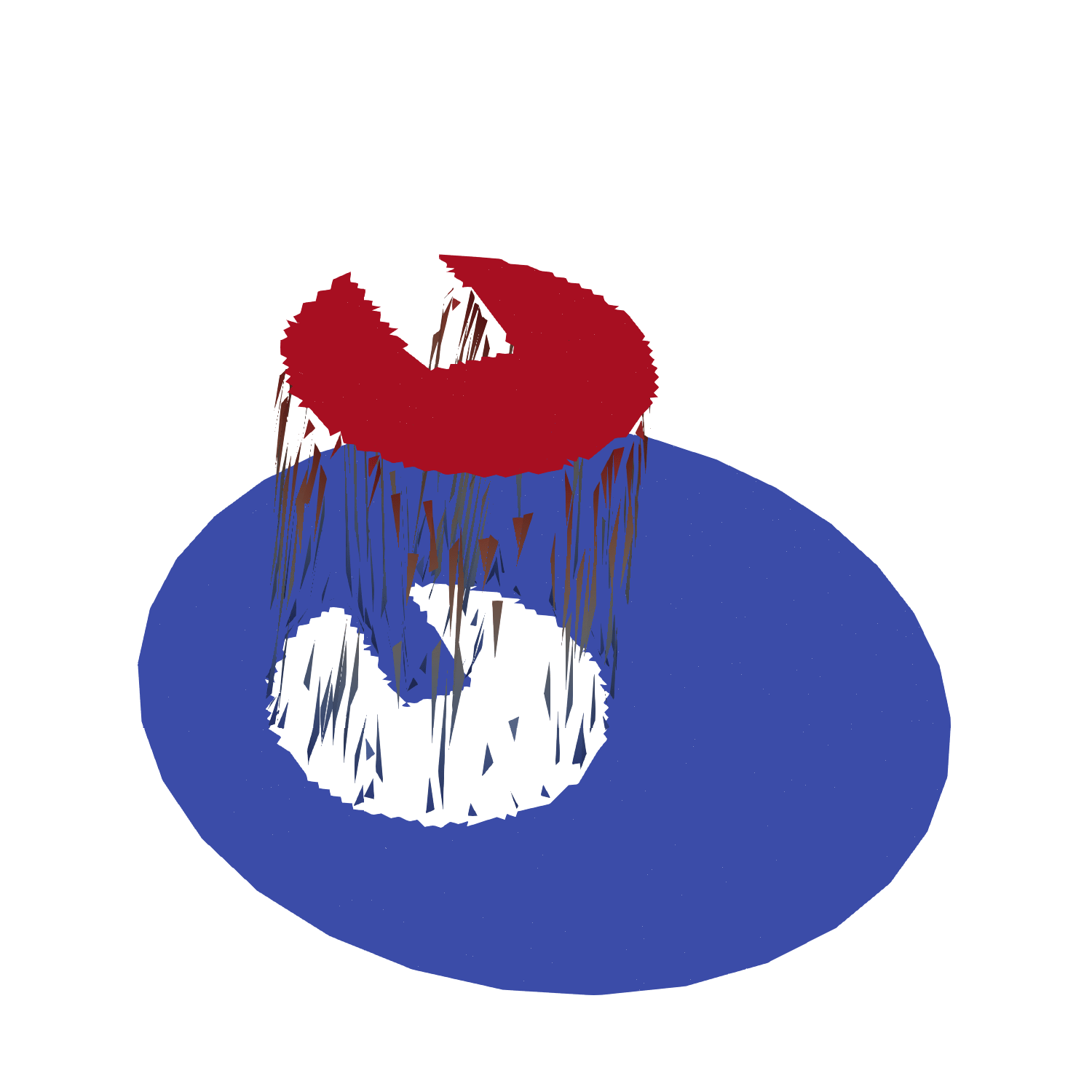}
        \caption{Case~3, $t=2$.}
    \end{subfigure}
    \caption{Slotted Disk: reconstructed mesh field $\psi_h$ at different time instants using a particle-mesh projection based on a bounded $\ell^2$-projection (Case~1), or a  PDE-constrained projection with $\zeta=0$ (Case~2) and $\zeta=30$ (Case~3).}
    \label{fig:slotted_disk_reconstructed_field}
\end{figure}

Fig.~\ref{fig:slotted_disk_reconstructed_field} compares the reconstructed mesh fields for the three different cases at time instants $t=1$ and $t=2$. For the bounded $\ell^2$-projection (Case~1), the discontinuity in the particle field is captured at the mesh without under- or overshoot, and values stay within the prescribed bounds $0 
\leq \xDiscreteScalar{\psi} \leq 
1 $ to machine precision, see Table~\ref{tab:slotted_disk_numeric_results}. Another advantage of this approach is that it is fast, and easy to implement in parallel. However, as reported in Table~\ref{tab:slotted_disk_numeric_results}, the mass error for the bounded $\ell^2$-projection is non-zero. The latter issue is overcome by using the conservative PDE-constrained particle-mesh projection for the reconstruction of the mesh fields, Case~2 and Case~3. However, for Case~2~-~in which $\zeta=0$~-~localized over- and undershoot is observed near the discontinuities. As argued in \cite{Maljaars2019}, this artifact is a resolution issue with the mesh being too coarse to capture the sharp discontinuity at the particle level monotonically.

By setting $\zeta$ to a value of 30, i.e. approximately equal to the number of particles per cell, this issue is effectively mitigated, see also the minimum and the maximum values for $\psi_h$ over the entire computation reported in Table~\ref{tab:slotted_disk_numeric_results}. Table~\ref{tab:slotted_disk_numeric_results} also confirms global mass conservation for Case~2 and Case~3, local conservation properties for similar examples were demonstrated in \cite{Maljaars2019, Maljaars2019_LNCSE} and not further shown here. 
\begin{table}[H]
    \centering
    \small{
    \begin{tabular}{c|c c c | c c c}
        \hline 
        \hline 
                &   Cells   & Particles & Wallclock time (s) & $\epsilon_{\Delta \psi_\Omega } (t=2)$ & $\psi_{h,\min}(x, t)$ & $\psi_{h,\max}(x, t)$ \\
        \hline 
         Case~1 &   14,464   &  438,495 & 150   &  1.4e-5   & -2.6e-16  & 1.00\\
         Case~2 &   14,464   &  438,495 & 182   &  2.0e-15  & -11.2 & 8.04\\
         Case~3 &   14,464   &  438,495 & 180   &  1.3e-15  & -0.01 & 1.02 \\
         \hline 
         \hline 
    \end{tabular}
    }
    \caption{Slotted Disk: runtime, area error $\epsilon_{\Delta \psi_\Omega }$ at $t=2$ and minimum and maximum values for $\psi_h$ for a bounded $\ell^2$-projection (Case~1), or a  PDE-constrained projection with $\zeta=0$ (Case~2) and $\zeta=30$ (Case~3).}
    \label{tab:slotted_disk_numeric_results}
\end{table}

\subsection{Lock exchange test}
As an example which is geared towards practical multi-fluid applications, the lock-exchange test is considered. Driven by gravity, a dense fluid current moves underneath a lighter fluid, where a thin vertical membrane initially separates the two fluids. Using \leopart, density tracking and momentum advection is done using Lagrangian particles, and the incompressible, unsteady Stokes equations are discretized on the mesh using the hybridized discontinuous Galerkin (HDG) method from \cite{Rhebergen2017, Rhebergen2018}. The computer code for this test can be found in \href{https://bitbucket.org/jakob_maljaars/leopart/src/master/tests/two_fluids/LockExchange.py}{\texttt{LockExchange.py}}. 

For this test, the density ratio $\gamma$ between the two fluids is 0.92. Furthermore, the domain of interest is $ \Omega := \left[0, 30 \right] \times \left[-0.5, 0.5\right] $, which is triangulated into $ 2,000 \times 80 \times 2 = 320,000$ regular triangular cells. Using the HDG method \cite{Rhebergen2018} with $ k=1 $ for solving the Stokes equations, the number of dofs in the global system amounts to 1,288,322. A direct sparse matrix solver (\texttt{SuperLU}) is used to solve the Stokes system. The total number of particles amounts to 9,408,000, where this number stays constant throughout the computation. Furthermore, 800 time steps of size $ \Delta t^{*} = \Delta t \sqrt{g'/H} = \expnumber{1.25}{-2}$ are performed, in which $H =1$ the channel height, and 
$ g' = g(1-\gamma) $ the reduced gravity with $\gamma=0.92$ and $g=9.81$.  

\begin{figure}[h]
\centering
\begin{subfigure}{\textwidth}
	\includegraphics[width=\textwidth]{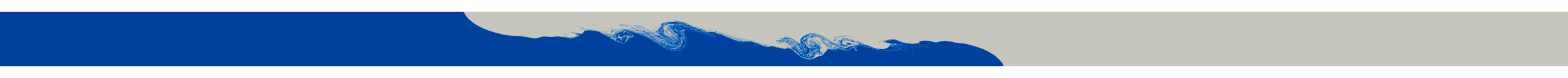}
	\caption{$ \ell^2 $-projections}
\end{subfigure}
\vspace{-10pt}
\begin{subfigure}{\textwidth}
	\includegraphics[width=\textwidth]{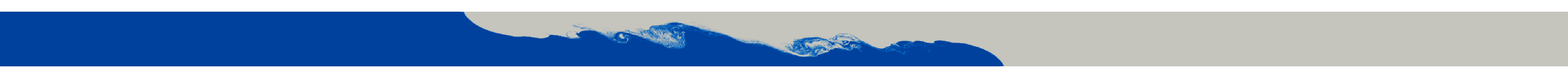}
	\caption{PDE-constrained projections}
\end{subfigure}
\caption{Lock exchange: density field at the mesh at $ t^* = 10 $ using $ \ell^2 $ or PDE constrained particle-mesh projections.}
\label{fig:lock_exchange_fields}
\end{figure}

Two different particle-mesh projection configurations are considered. In a first configuration, the local $ \ell^2 $-projections are used to project density and momentum data from the particles to the mesh. The density projection is rendered bound preserving by imposing box constraints on the local least squares problem. The density values attached to a particle stay constant throughout the computation, whereas the specific momentum value attached to a particle is updated using Eq.~\eqref{eq:discrete_particle_update} after the Stokes solve. 

In the second configuration, the PDE-constrained projection is used for the projection of density and momentum data from the particles to the mesh. This results in a global problem with 964,160 unknowns for the density projection, and 1,928,320 unknowns for the momentum projection. The global systems resulting from the PDE-constrained projections are solved using a GMRES solver in conjunction with an algebraic multigrid preconditioner, where this solver/preconditioner pair is used as a black-box. 
Furthermore, for the density projection, $\zeta=20$ to penalize over- and undershoot in the reconstructed density fields, and as for the other configuration, the specific momentum value attached to a particle is updated using Eq.~\eqref{eq:discrete_particle_update}.

\begin{figure}[H]
    \centering
    \begin{subfigure}{\textwidth}
        \centering
    	\includegraphics{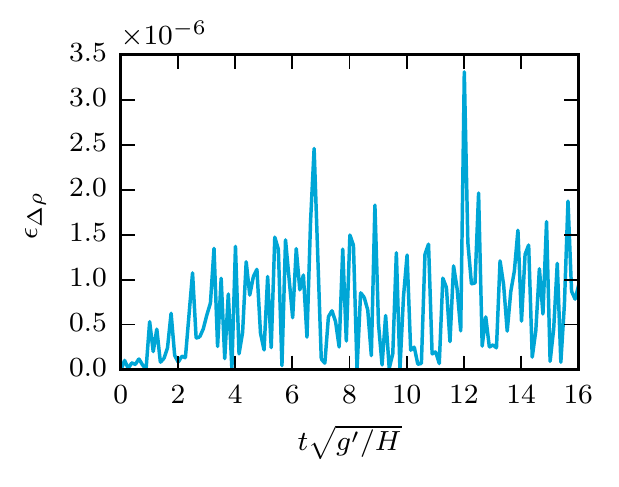}
    	\includegraphics{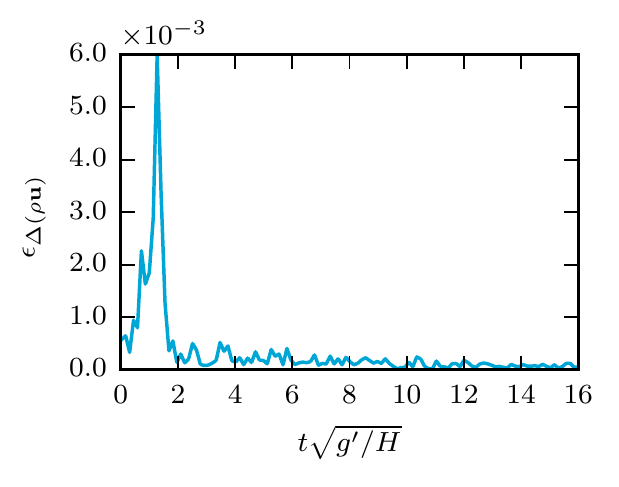}
    	\caption{$ \ell^2 $-projection: mass- (left) and momentum error (right).}
    \end{subfigure}
    \begin{subfigure}{\textwidth}
        \centering
    	\includegraphics{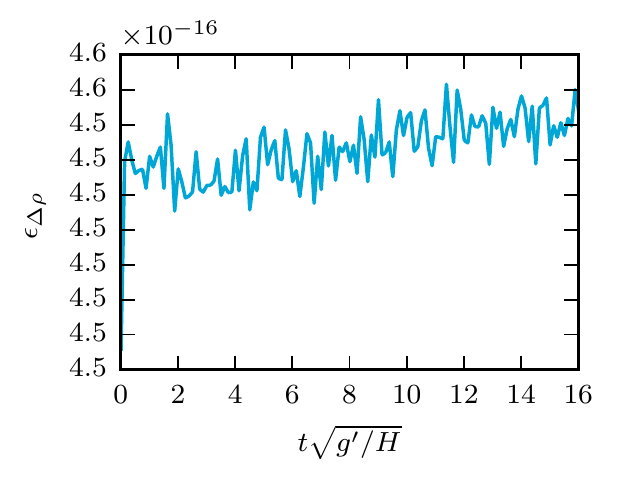}
    	\includegraphics{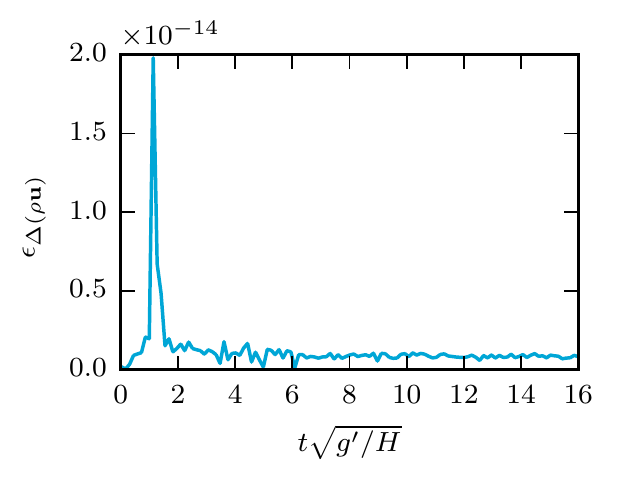}
    	\caption{PDE-constrained projection: mass- (left) and momentum error (right).}
    \end{subfigure}
    \caption{Lock exchange: mass- and momentum conservation error as a funtion of time for two different particle-mesh projection strategies.}
    \label{fig:lock_exchange_mass_momentum}
\end{figure}

Visually, the mesh density fields at $ t^* = 10 $ are comparable in terms of the bulk behavior for both projections, Fig.~\ref{fig:lock_exchange_fields}, although differences in the small scale features are observed. No further attempts are made to interpret and value these small scale differences between the local $ \ell^2 $-projections and the PDE-constrained projections, other than to say that the PDE-constrained approach results in mass- and momentum-conservative fields, whereas this is not so for the $\ell^2$-projection, see Fig.~\ref{fig:lock_exchange_mass_momentum}.

\begin{figure}[H]
\centering
\begin{subfigure}{0.495\textwidth}
	\includegraphics{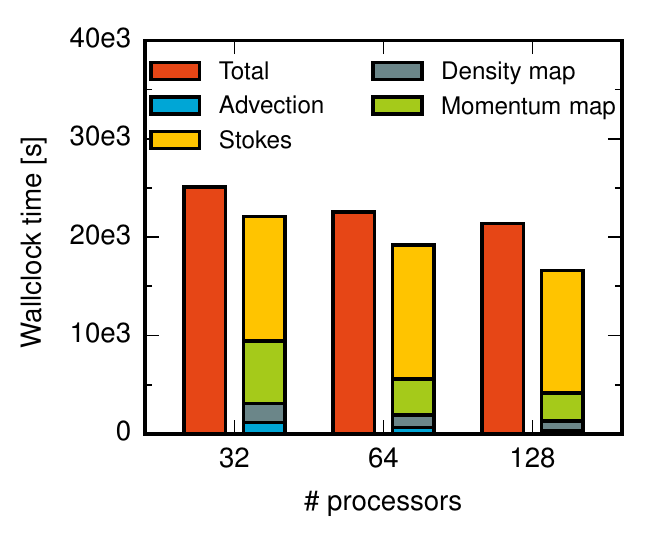}
	\caption{PDE-constrained projections}
	\label{fig:le_PDE_wallclock}
\end{subfigure}
\begin{subfigure}{0.495\textwidth}
	\includegraphics{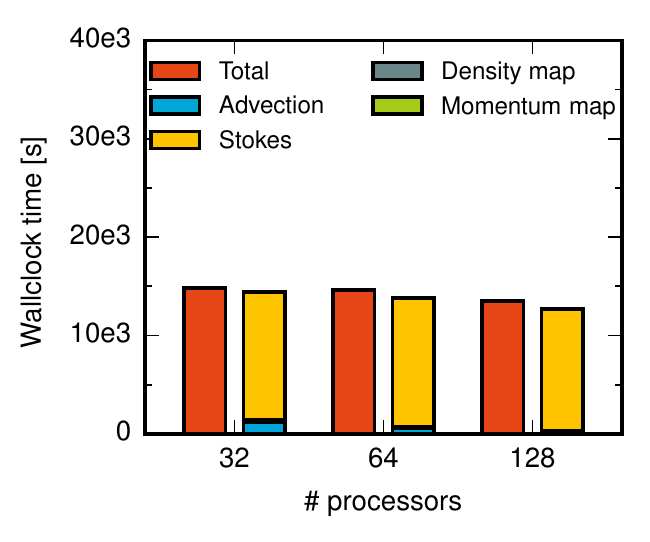}
	\caption{$ \ell^2 $-projections}
	\label{fig:le_l2_wallclock}
\end{subfigure}
\caption{Lock exchange: strong scaling study.}
\label{fig:lockexchange_wallclock}
\end{figure}

Timings are reported in Fig.~\ref{fig:lockexchange_wallclock}, using 32, 64 and 128 CPU cores on the Peta~4 supercomputing facility of the University of Cambridge. 
Peta~4 contains 768 nodes, equipped with $ 2 $ Intel Xeon Skylake 6142 16-core processors each. 

Results provide insight into the performance of the different parts, and indicate which parts of the scheme are critical for obtaining higher performance. Clearly, the computational time is dominated by the global solves for the Stokes system, and the PDE-constrained particle-mesh projections, Fig.~\ref{fig:le_PDE_wallclock}. The advantage of using iterative solvers for the PDE-projections also becomes clear from this figure. Even though the system for the momentum projection is larger than the system for the Stokes projection, the wallclock time for the momentum projection is considerably smaller and appears to possess better scaling compared to the Stokes solve. Therefore, implementing the iterative solver for the Stokes solver proposed in \cite{Rhebergen2018_2} is believed to be an important step for improving the performance, and probably indispensable for problems in three spatial dimensions. Noteworthy to mention is that the $ \ell^2 $ particle-mesh projections exhibit excellent scaling properties, on top of their low computational footprint, see Fig.~\ref{fig:le_l2_wallclock}.

\begin{table}[H]
\centering
\caption{Lock exchange test: speed-up of the different model parts in parallel computations, benchmarked against 32 processors.}
\label{tab:speed-up lock exchange}
\small{
\begin{tabular}{c | c c c | c c c}
\hline
\hline
& \multicolumn{3}{c|}{PDE-projections}	& \multicolumn{3}{c}{$ \ell^2 $ projections}	\\
\hline
& \multicolumn{3}{c|}{\# Processors}  & \multicolumn{3}{c}{\# Processors}			\\
							& 32 	&	64	& 	128  &	32	& 64 	& 128 		\\
\hline
Particle advection			& 1		& 1.85   & 	3.87 &   1   & 2.06	& 3.95		\\
Density projection			& 1		& 1.48   & 	1.87 &   1   & 1.75	& 3.75		\\
Momentum projection			& 1		& 1.73	 & 	2.25 & 	1   & 1.79  & 3.99		\\	
Stokes solve				& 1 	& 0.93	 &  1.17 &	1	& 0.99 	& 1.05		\\
\hline
Total						& 1		& 1.11	 & 	1.17 & 	1	& 1.01	& 1.1		\\
\hline
\hline
\end{tabular}
}
\end{table}

\subsection{Rayleigh--Taylor instability benchmark}
In this example, the applicability of \leopart\ for simulating active particle tracing problems is demonstrated. A well-established benchmark from the geodynamics community is used to fit this purpose, and we consider the Rayleigh-Taylor instability
community benchmark proposed by van Keken and co-workers~\cite{PvK1997}. This benchmark involves the evolution of a geodynamic laminar flow
whose initial state is a compositionally light material, situated under a 
considerably thicker and denser layer. Tools from \leopart\ are used to discretise the Stokes system using the HDG method, and Lagrangian particles are used for a diffusion-free tracking of the chemical composition field. Conservative composition fields at the mesh are reconstructed from the particle representation using the PDE--constrained projection.
The reconstructed composition field is subsequently used to compute a source term $\mathbf{f}$ and a composition-dependent viscosity $\eta$ in the Stokes equations.
The code for this numerical experiment
can be found in \href{https://bitbucket.org/jakob_maljaars/leopart/src/master/tests/two_fluids/RayleighTaylorInstability.py}{\texttt{RayleighTaylorInstability.py}}, 
considering the following problem description:



Let the domain $\Omega$ be the $[0, L] \times [0, 1]$ rectangle, where $L=0.9142$
is the aspect ratio. $\psi : \Omega \rightarrow [0, 1]$ is the continuum representation of the
chemical composition function, with values $0$ and $1$ corresponding to the 
light and dense layer. 
The source term in the momentum component of Stokes' equations is given by
\begin{equation}
    \mathbf{f} = \mathrm{Rb} \, \psi \, \hat{\mathbf{g}},
    \label{eq:rt_momentum_source}
\end{equation}
where $\mathrm{Rb} = 1$ is the compositional Rayleigh number and
$\hat{\mathbf{g}} = (0, -1)^\top$ is the unit vector acting in the direction
of gravity. 
%
%
The viscosity of the Stokes system is dependent on the chemical
composition function
%
\begin{equation}
    \eta = \eta_\text{bottom} + \psi ( \eta_\text{top} - \eta_\text{bottom} ),
    \label{eq:rt_viscosity_model}
\end{equation}
%
where $\eta_\text{top}$ and $\eta_\text{bottom}$ are the viscosities of
the initial heavy top and light bottom layers, respectively. The initial state of $\psi$
is a small perturbation from the rest state of a light layer of depth
$d_b = 0.2$,
%
\begin{equation}
    \psi(\mathbf{x}, t=0) = 
   \begin{cases} 
      0 & y < d_b + 0.02 \cos \left( \frac{\pi x}{L} \right),  \\
      1 & \text{otherwise}.
   \end{cases}
   \label{eq:initial_eta}
\end{equation}
%
The boundary conditions are set such that $\mathbf{u} = \mathbf{0}$ on the bottom ($y=0$) and top ($y=1$) boundaries, and on the left ($x=0$) and right ($x=L$) boundaries, freeslip conditions are applied, i.e. $\mathbf{u}\cdot \unitnormal = 0$ and $\unittangent\cdot \left(\nabla \mathbf{u} + \nabla \mathbf{u}^\top \right)\unitnormal = \mathbf{0}$ where $\unitnormal$ the outward pointing unit normal vector and $\unittangent$ is the unit tangent vector to the boundary. 
Furthermore, the boundary conditions imposed on the chemical composition function are $\psi = 1$ on top ($y=1$) and $\psi = 0$ on the bottom ($y=0$) boundaries,
respectively.

The domain is triangulated into $2n^2$ regular simplices, with $n=(40,\, 80,\, 160)$, yielding three meshes with 3200, 12800, and 51200 cells respectively.
%
%
The mesh is then displaced in order to align the cells with the
initial viscosity discontinuity described in~\eqref{eq:initial_eta}. 
Each cell is assigned \num{25}~particles, carrying a composition value $\psi_p$, resulting in \num{80000}, \num{320000} and 
\num{1280000}~particles in the meshes, respectively. This number of particles remains 
constant throughout the simulation.
Furthermore, the time
step size is chosen based on the Courant--Friedrichs--Levy (CFL) criterion, i.e.
$\Delta t_j = {C_\mathrm{CFL} h_\text{min}} / {\max_\Omega |\mathbf{u}(t_j)|}$ 
where $t_j$ is the $j$th time step, $h_\text{min}$ is the minimum mesh cell diameter
and $C_\mathrm{CFL} > 0$ is the CFL parameter, chosen here to be $C_\mathrm{CFL} = 0.5$.
The HDG method is used with $k=1$ to solve the Stokes system.  
After computing
the solution approximation of the Stokes system, the particles are advected. Given $\psi_p$, the conservative PDE--constrained projection is used to update the composition field $\psi_h$ on the mesh and thereby the source term $\mathbf{f}$ and the viscosity $\eta$.
The composition function $\psi_h$ is represented by the $k=1$ DG--finite element space,
and we choose
$\zeta = \num{25}$ to penalise over-- and undershoot of the the reconstructed field.

\begin{figure}[H]
\centering
    \begin{subfigure}{0.275\textwidth}
    	\centering
        \includegraphics[width=0.99\linewidth]{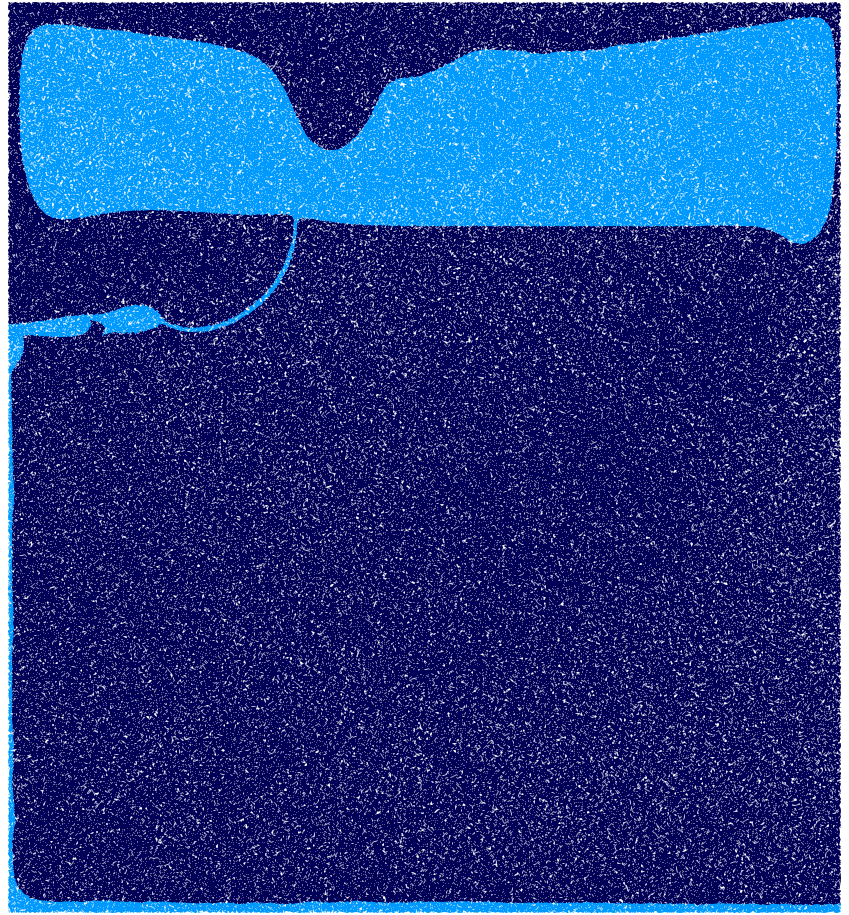}
        \caption{$t = \num{500}$}
        \label{fig:rayleightaylor_functionals:urms}
    \end{subfigure}
    \hfill
    \begin{subfigure}{0.275\textwidth}
        \centering
        \includegraphics[width=0.99\linewidth]{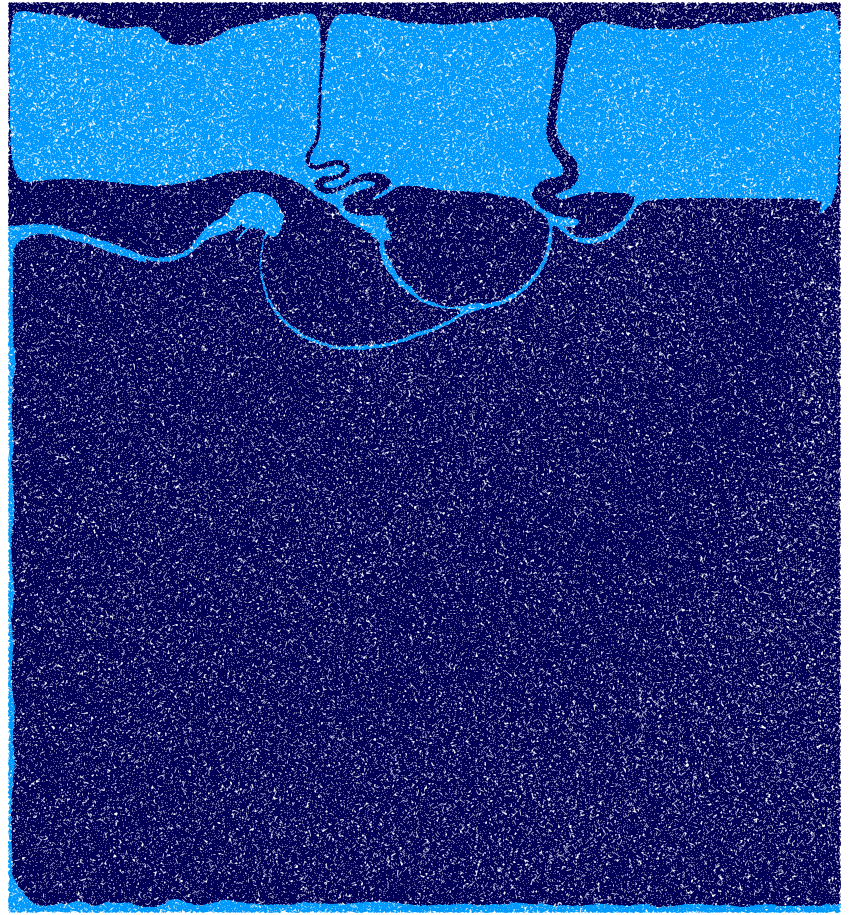}
        \caption{$t = \num{1000}$}
    \end{subfigure}
    \hfill
    \begin{subfigure}{0.275\textwidth}
        \centering
        \includegraphics[width=0.99\linewidth]{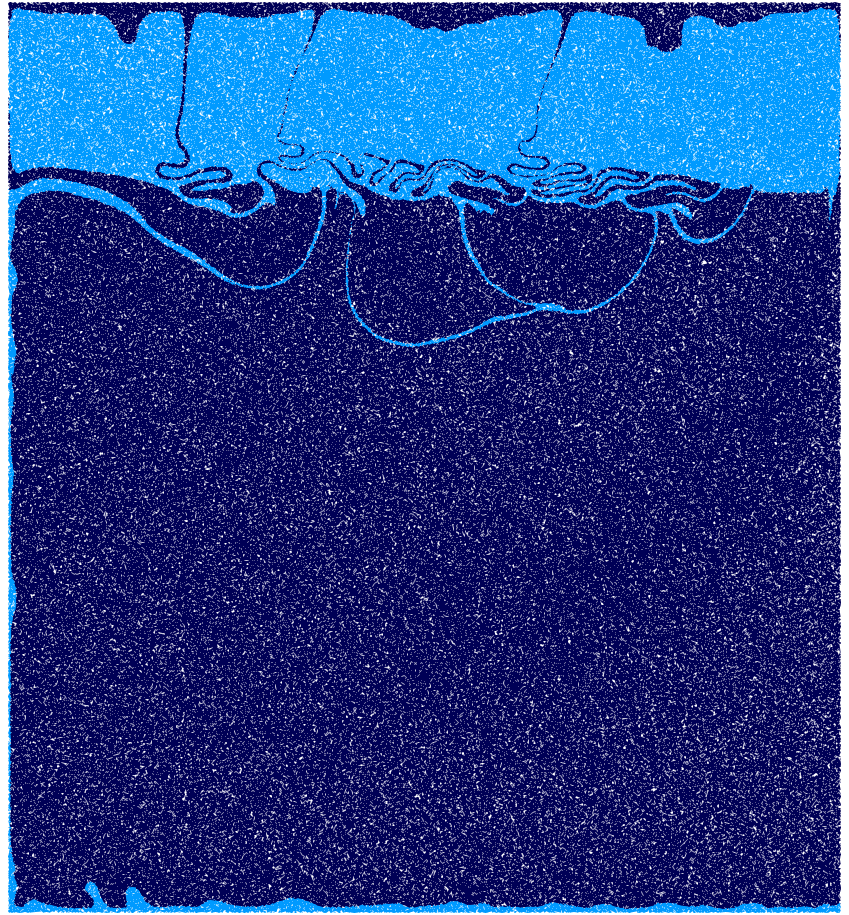}
        \caption{$t=\num{1500}$}
    \end{subfigure}
    \caption{Rayleigh Taylor instability: the \num{1280000} particles distributed on the fine resolution mesh with $n=160$ at given simulation times. The dark and light colours represent compositionally dense $\psi = 1$ and light $\psi = 0$ regions, respectively. Cf.~\cite{PvK1997}.}
    \label{fig:rti_particle_plot}
\end{figure}

The benchmark~\cite{PvK1997} considers three cases, $\eta_\text{top} / \eta_\text{bottom} \in
\{1, 10, 100\}$. For brevity we document the case with a \num{100}--fold difference 
in the viscosity layers, namely, $\eta_\text{top} = 1$ and
$\eta_\text{bottom} = 0.01$. For comparison with~\cite{PvK1997}, the distribution of the \num{1280000} particles in the 
%
%
$n=160$
mesh
at simulation times $t = \num{500}, \num{1000}$ and $\num{1500}$ are shown in Fig.~\ref{fig:rti_particle_plot}. Noteworthy to mention is that the particle distribution remains uniform throughout. This feature owes to HDG discretization of the Stokes system, yielding pointwise div-free velocity fields by which the particles are advected \cite{Maljaars2017}.

\begin{figure}[H]
\centering
    \begin{subfigure}{0.45\textwidth}
    	\centering
        \includegraphics[width=0.99\linewidth]{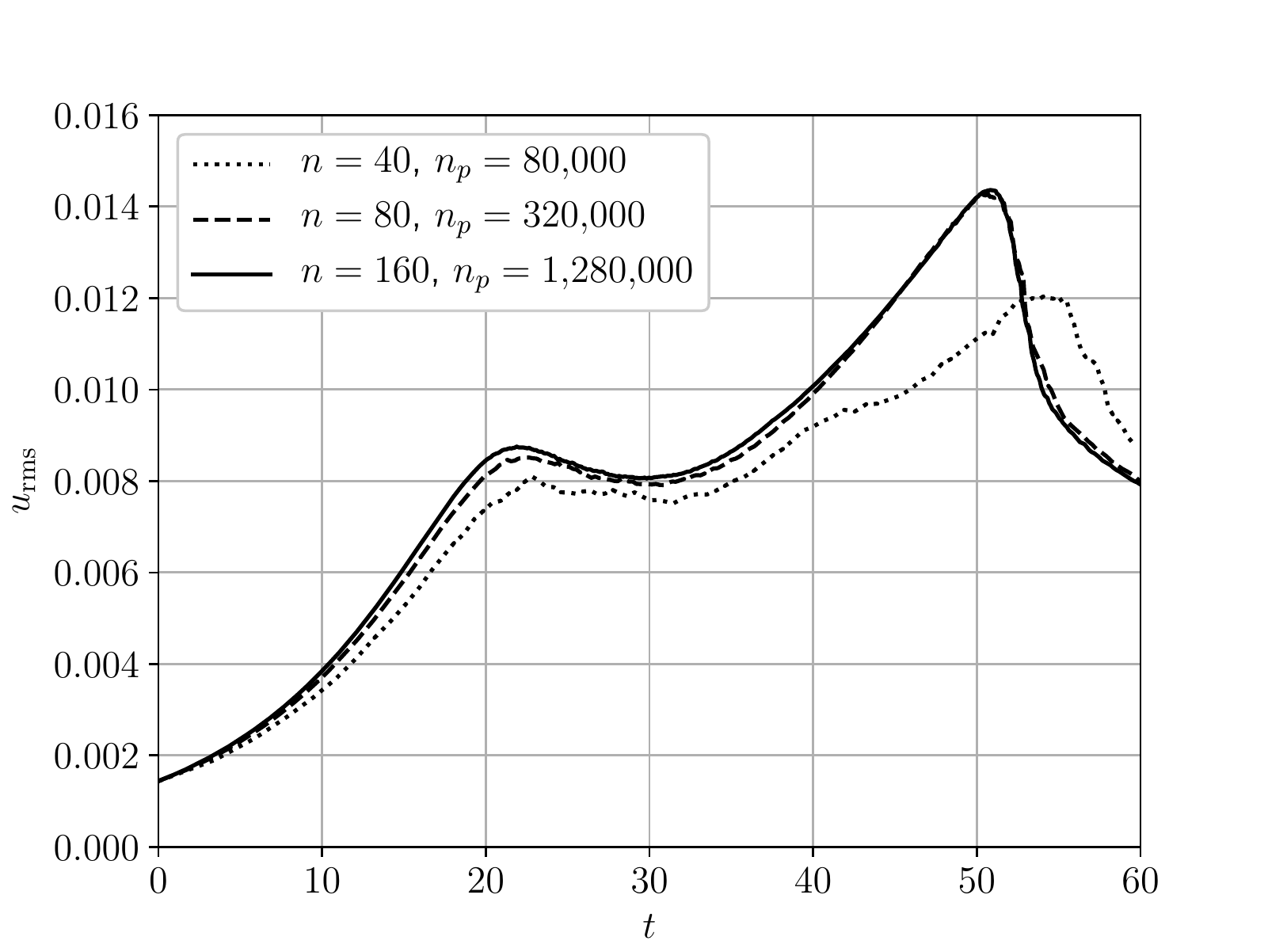}
        \caption{RMS velocity}
        \label{fig:rti_functionals:urms}
    \end{subfigure}
    \begin{subfigure}{0.45\textwidth}
        \centering
        \includegraphics[width=0.99\linewidth]{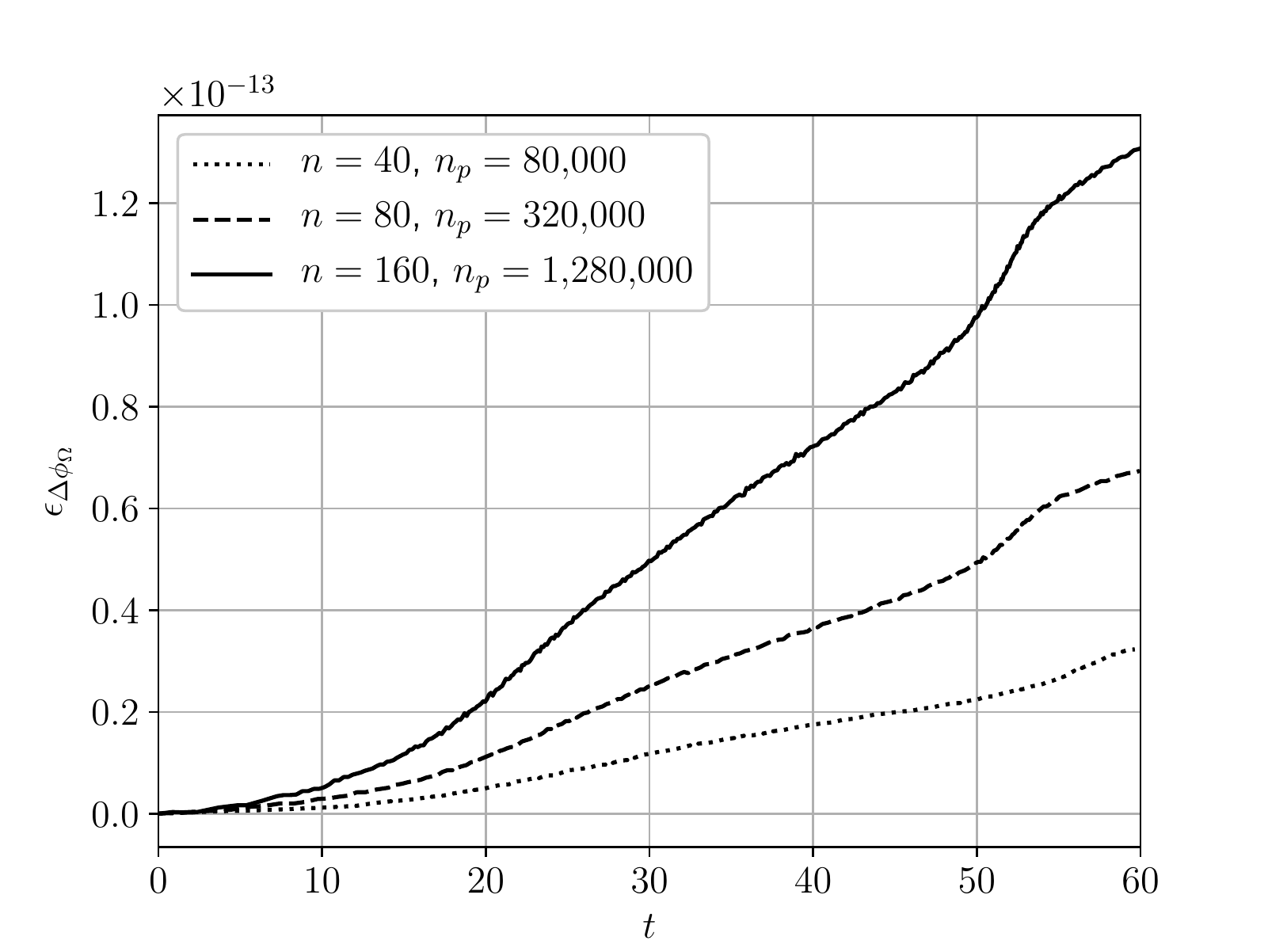}
        \caption{Mass conservation error}
        \label{fig:rti_mass_conn}
    \end{subfigure}
    \caption{The computed RMS velocity functional and mass conservation error
    of the Rayleigh Taylor instability problem. Here $n$ is the number of cells along
    one axis of the mesh and $n_p$ is the number of particles used in the simulation.}
    \label{fig:rti_functionals}
\end{figure}

A quantitative comparison with literature results from \cite{PvK1997, Vynnytska2013} is made by investigating 
the root mean square (RMS) velocity, $u_\mathrm{rms}$, the mass conservation error, $\epsilon_{\Delta\psi_\Omega}$, and growth rate, $\gamma$, where
%
\begin{align}
u_\mathrm{rms} = \sqrt{\frac{\areaIntegral{\Omega}{\mathbf{u} \cdot \mathbf{u}}}{\areaIntegral{\Omega}{}}}, &&
\epsilon_{\Delta\psi_\Omega} = \left| \areaIntegral{\Omega}{\left( \psi_h(\mathbf{x}, t) - \psi_h(\mathbf{x}, 0) \right)} \right|, &&
\gamma = \frac{\ln \left(u_\mathrm{rms}(t_1)\right) - \ln \left(u_\mathrm{rms}(0)\right)}{t_1},
\label{eq:rti_functionals}
\end{align}
%
and $t_1$ is the simulation time at the first time step.
The computed quantities~\eqref{eq:rti_functionals}
are shown in Fig.~\ref{fig:rti_functionals}, and key results are
reported in Table~\ref{tab:rti_results}. This includes comparison to the
marker chain method of~\cite{PvK1997}
and the case with no artificial diffusion in~\cite{Vynnytska2013}, showing good agreement for the growth rate $\gamma$ and the RMS-velocity. On top of this, discrete conservation for the composition field $\psi_h$ can be ensured, Fig.~\ref{fig:rti_mass_conn}, when using the PDE-constrained particle-mesh projection provided by \leopart.

\begin{table}[H]
\centering
\small{
\begin{tabular}{lrl|rrr} \hline \hline
Code             & Grid             & Method             & $\gamma$ & $t(\max u_\mathrm{rms})$ & $\max u_\mathrm{rms}$ \\ \hline 
van~Keken~et~al.~\cite{PvK1997} & $100 \times 100$ & splines / marker chain            & 0.1024               & 51.12                    & 0.01385               \\
                 & $80 \times 80$   & $C1$--element / marker chain     & 0.1025               & 51.23                    & 0.01448               \\ \hline
Vynnytska~et~al.~\cite{Vynnytska2013} & $40 \times 40$   & Taylor--Hood / DG  & 0.0976               & 57.21                    & 0.01140               \\
                 & $80 \times 80$   & Taylor--Hood / DG  & 0.1018               & 52.11                    & 0.01444               \\
                 & $160 \times 160$ & Taylor--Hood / DG  & 0.1039               & 51.55                    & 0.01458               \\ \hline
This work        & $40 \times 40$   & HDG / PDE--constrained projection & 0.08707              & 54.81                    & 0.01208               \\
                 & $80 \times 80$   & HDG / PDE--constrained projection & 0.09161              & 50.68                    & 0.01428               \\
                 & $160 \times 160$ & HDG / PDE--constrained projection & 0.09677              & 50.80                    & 0.01436             \\
                 \hline \hline
\end{tabular}
}
\caption{Computed functionals from the Rayleigh Taylor instability simulation and comparison
with other works.}
\label{tab:rti_results}
\end{table}

\subsection{Rayleigh--Taylor instability -- 3D}
We finally demonstrate the use of \leopart\ for 3D simulations. The 3D example in this section is constructed by
extruding the Rayleigh--Taylor instability benchmark from the previous section. Let the domain $\Omega := [0, L_x] \times [0, 1] \times [0, L_z]$
where $L_x = \num{0.9142}$ and $L_z = \num{0.8142}$. The momentum source and viscosity model is as described in~\eqref{eq:rt_momentum_source}
and~\eqref{eq:rt_viscosity_model}, respectively. The initial perturbed composition field is

\begin{equation}
    \psi(\mathbf{x}, t=0) = 
  \begin{cases} 
      0 & y < d_b + 0.02 \cos \left( \frac{\pi x}{L_x} \right) \cos \left( \frac{\pi z}{L_z} \right),  \\
      1 & \text{otherwise}.
  \end{cases}
  \label{eq:initial_eta_3d}
\end{equation}
%
The boundary conditions are imposed as described in the previous section with the addition of a free slip boundary condition enforced on the near ($z=0$) and far ($z=L_z$) boundaries.  All other parameters are as described in the previous section. The code for this example can be found in \href{https://bitbucket.org/jakob_maljaars/leopart/src/master/tests/two_fluids/RayleighTaylorInstability3D.py}{\texttt{RayleighTaylorInstability3D.py}}

The domain is divided into $6n^3$ tetrahedra where $n = 20$. 
The mesh is displaced to align with the viscosity discontinuity in Eq.~\eqref{eq:initial_eta_3d}.
Each cell is assigned \num{50}~particles, carrying a composition value $\psi_p$, such that there is a 
total of \num{2400000}~particles used in the simulation.

Evidently Fig.~\ref{fig:rti_functionals:urms} shows that high resolution meshes are mandatory
for accurate results. To solve large 3D problems efficiently we refer to an example of a preconditioner 
designed for the iterative solution of the HDG system~\cite{Rhebergen2018_2}.
Implementation of this preconditioned system forms a programme of future development in \leopart. 
In this example we use the direct solver MUMPS to solve the underlying linear system.

\begin{figure}[H]
\centering
    \begin{subfigure}{0.3\textwidth}
    	\centering
        \includegraphics[width=0.99\linewidth]{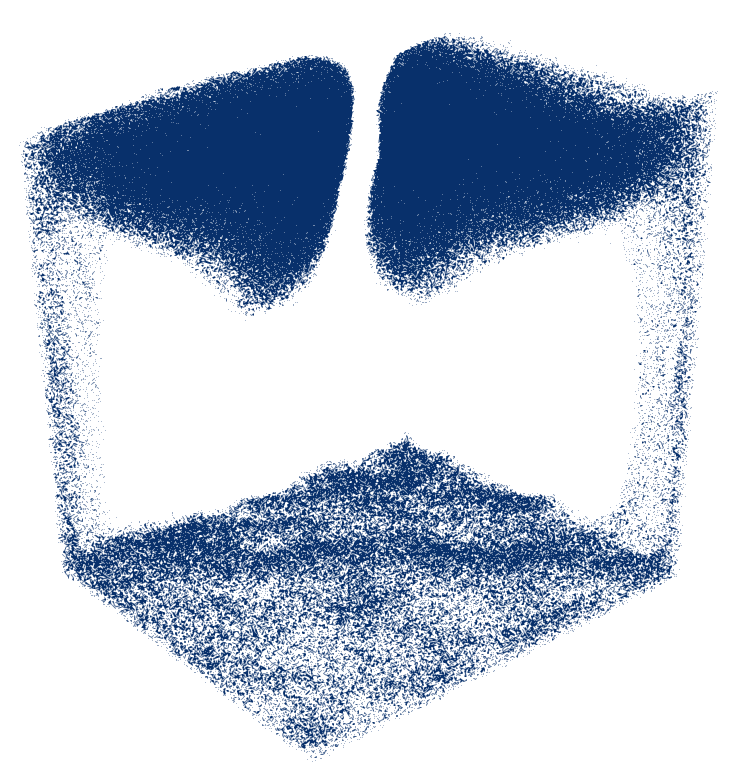}
        \caption{$t = \num{250}$}
    \end{subfigure}
    \hfill
    \begin{subfigure}{0.3\textwidth}
        \centering
        \includegraphics[width=0.99\linewidth]{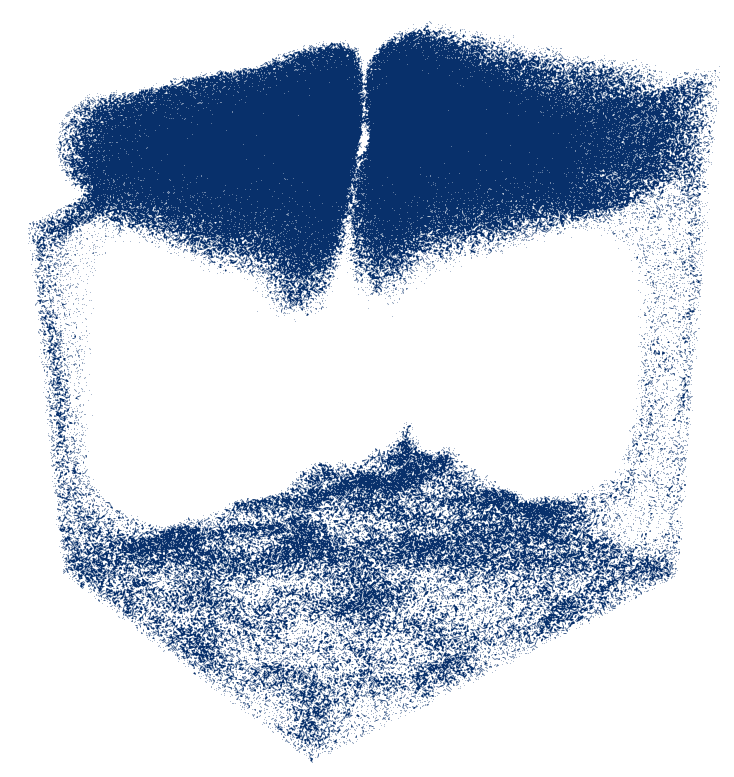}
        \caption{$t = \num{500}$}
    \end{subfigure}
    \hfill
    \begin{subfigure}{0.3\textwidth}
        \centering
        \includegraphics[width=0.99\linewidth]{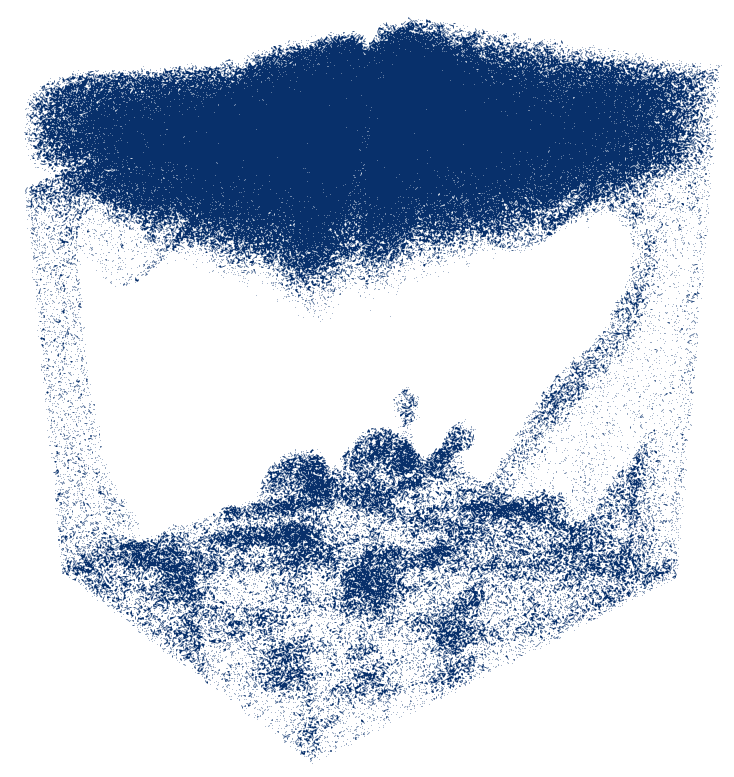}
        \caption{$t=\num{1000}$}
    \end{subfigure}
    %
    \begin{subfigure}{0.05\textwidth}
        \centering
        \includegraphics[width=0.99\linewidth]{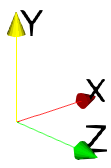}
    \end{subfigure}
    \caption{3D Rayleigh Taylor instability: the compositionally light ($\psi = 0$) particles distributed on the mesh where $n=20$ at given simulation times.}
    \label{fig:rti_particle_plot_3d}
\end{figure}

The tracer distribution of the compositionally light ($\psi = 0$) material is shown in 
Fig.~\ref{fig:rti_particle_plot_3d}. We see the formation of two opposing diapirs competing for
space at the top of the domain. The interface separating the two diapirs is formed by
a downwelling of compositionally dense material ($\psi = 1$) from the top of the domain.

\begin{figure}[H]
\centering
    \begin{subfigure}{0.45\textwidth}
    	\centering
        \includegraphics[width=0.99\linewidth]{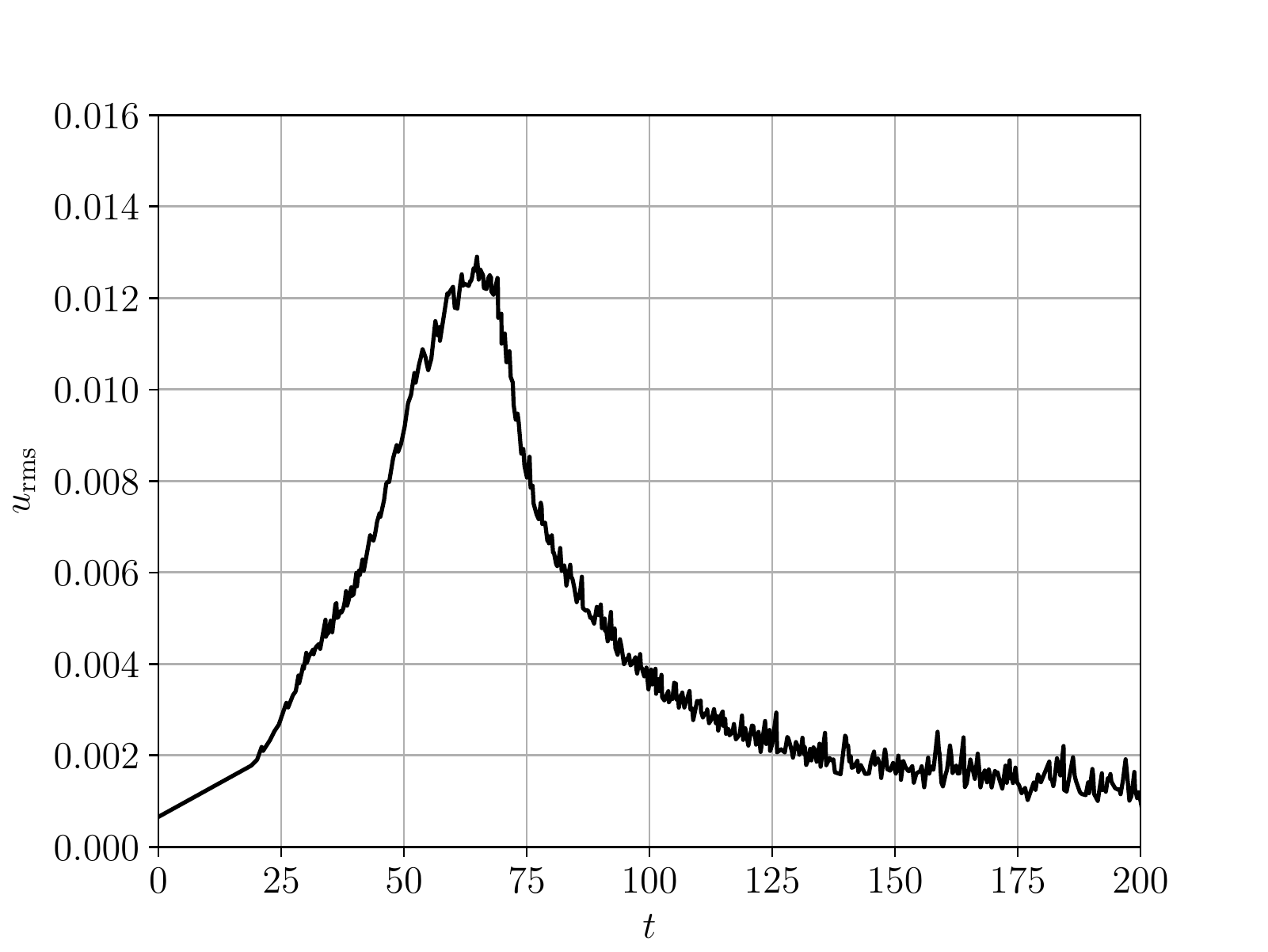}
        \caption{RMS velocity}
        \label{fig:rti_functionals_3d:urms}
    \end{subfigure}
    \begin{subfigure}{0.45\textwidth}
        \centering
        \includegraphics[width=0.99\linewidth]{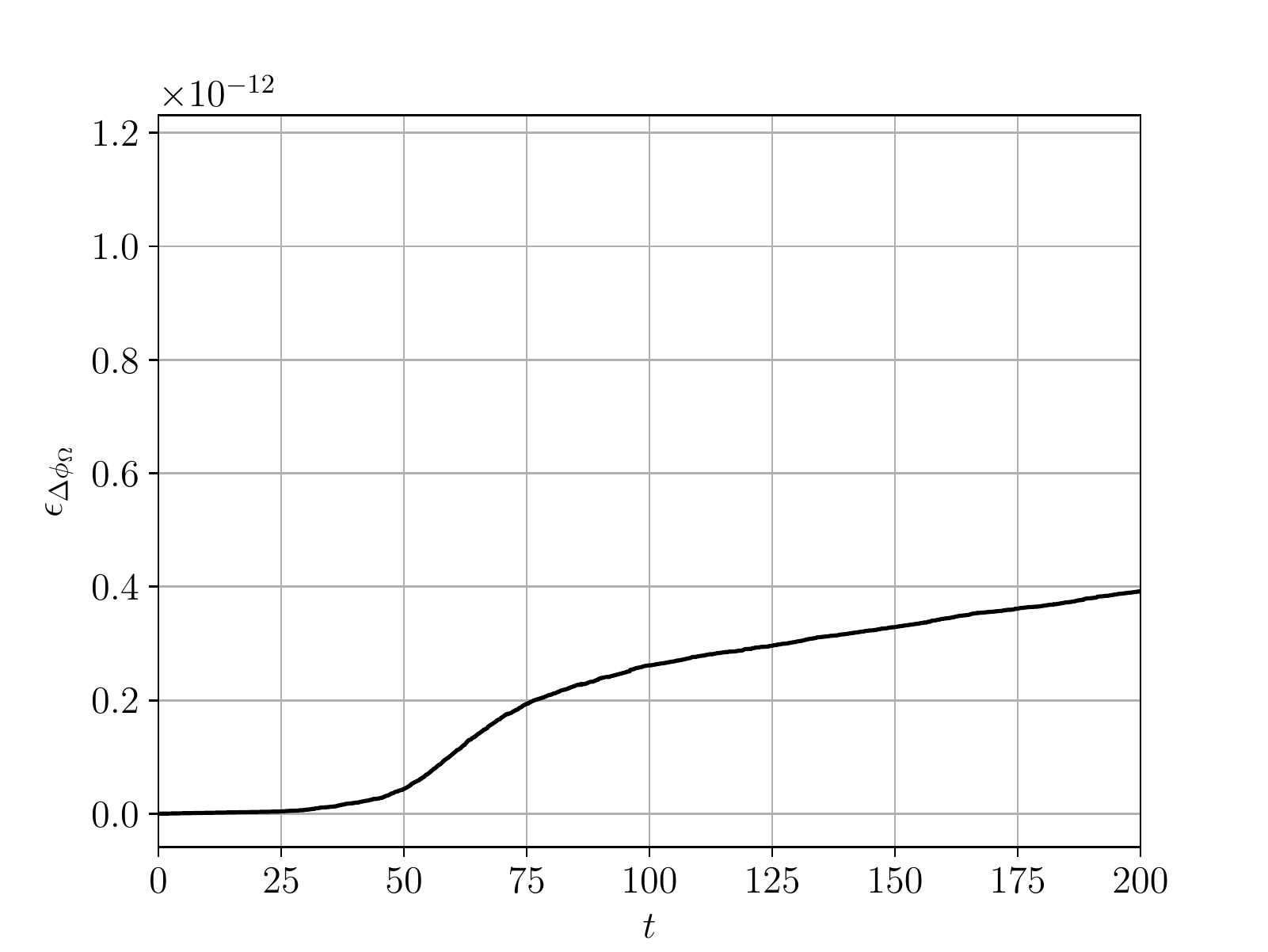}
        \caption{Mass conservation error}
        \label{fig:rti_mass_conn_3d}
    \end{subfigure}
    \caption{The computed RMS velocity functional and mass conservation error
    of the 3D Rayleigh Taylor instability problem. Here $n = \num{20}$ is the number of cells along
    one axis of the mesh and $n_p = \num{2400000}$ is the number of particles used in the simulation.}
    \label{fig:rti_functionals_3d}
\end{figure}

We show RMS velocity and mass conservation in Fig.~\ref{fig:rti_functionals_3d}. Here
we confirm the mass conserving property of the PDE--constrained projection of the
particle composition values to the composition function. Furthermore the two competing
diapirs evolve more slowly than the single diapir exhibited in the previous section.

\section{Conclusion and outlook} \label{sec:conclusions}
This paper introduced \leopart\ \cite{LEoPart}, an open-source library which integrates a suite of tools for Lagrangian particle advection and different particle-mesh interaction strategies in \fenics. To efficiently implement the particle advection, particles  are tracked on the simplicial mesh using the convex polyhedron method. As demonstrated in the numerical examples, the particle advection exhibits near optimal performance for distributed-memory parallel runs. Furthermore, several options are available for the projection of particle data onto mesh fields and \textit{vice versa}. In particular, the PDE-constrained particle-mesh projection allows to track particle quantities on the mesh in an accurate, diffusion-free, and conservative manner.

A number of application examples in two and three spatial dimensions demonstrated how \leopart\ can be used to track sharp interfaces in immiscible, multiphase flows or long time scale processes such as pertaining to geodynamics. Yet, we believe that the particle(-mesh) functionality in \leopart\ can be of practical relevance to a much wider range of flow problems, including groundwater modeling, atmospheric modeling, and reproducing particle image velocimetry measurements in numerical simulations. Of particular interest are applications characterized by low physical diffusion, for which the presented particle-mesh tools allow to maintain sharp features at subgrid level without introducing numerical diffusion. 

\leopart\ \cite{LEoPart} will be maintained at the cited URL, and the community is invited to contribute to this project. Upcoming developments in \leopart\ include:
\begin{itemize}[noitemsep]
    \item \textbf{Update to dolfin-x}: Recent developments in \fenics\ have been taking place in the dolfin-x repository, which has diverged significantly from the original dolfin. It is planned to migrate \leopart\ to the new underlying library. This will require many changes, including basic geometry handling and assembly of forms. However, we can expect to see some performance enhancements as a result, especially as it will be easier to assemble block diagonal matrices and preconditioners to solve the Stokes equations iteratively.
    \item \textbf{Iterative solvers:} as demonstrated in the numerical examples, all the components exhibit excellent scaling for distributed-memory parallel runs, except for solving the global systems which arise in the PDE-constrained projection and the incompressible Stokes equations. To improve the performance for these steps, the implementation of scalable iterative solvers heads our wish list. The optimal preconditioner presented by \cite{Rhebergen2018_2} will serve as a starting point for the Stokes system, whereas implementation of a GMRES-based solver is considered for the PDE-constrained projection.
    \item \textbf{Particle advection:} at the time of writing \leopart\ supports an explicit Euler, and a two- or three-stage Runge-Kutta scheme for the particle-advection. In view of particle-mesh operator splitting applications, \leopart\ will benefit from supporting multi-step schemes as this opens the way for implementing implicit-explicit (IMEX) operator splitting schemes \cite{Karniadakis1991}. Theoretically, such schemes can push particle-mesh operator splitting techniques beyond the second order time accuracy as reported in \cite{Maljaars2017, Maljaars2019}.
\end{itemize}

\section*{Acknowledgements}
The Netherlands Organisation for Scientific Research (NWO) is acknowledged for supporting JMM through the JMBC-EM Graduate Programme research grant. 
JMM further acknowledges CRUX Engineering for their time investment in the revision stage of the manuscript. 
CNR is supported by EPSRC Grant EP/N018877/1. NS is supported by the Carnegie Institution for Science President's Fellowship. NS further wishes to thank Peter E. van Keken and Cian R. Wilson for their advice.

\appendix
\section{PDE-constrained particle-mesh interaction}
\label{app:pde_constrained_pm}
This appendix presents the discrete optimality system which is obtained by equating the variations of Eq.~\eqref{eq:scalar_lagrangian-functional} with respect to the three unknowns $(\xDiscreteScalar{\psi}, \xDiscreteScalar{\lambda}, \xDiscreteScalar{\bar{\psi}}) \in (\xDiscreteScalar{W}, \xDiscreteScalar{T}, \bar{W}_{h})$ to zero, and performing a $\theta$ time integration. A detailed derivation can be found in \cite{Maljaars2019}.

To set the stage, let $\mathcal{T} := \{K\}$ be the triangulation of $\Omega$ into open, non-overlapping cells $K$, having outward pointing unit normal vector $\unitnormal$ on its boundary $\partial K$. Adjacent cells $K_i$ and $K_j$ ($i\neq j$) share a common facet $F =\partial{K}_i \cap \partial{K}_j$. The set of all facets (including the exterior boundary facets $F =\partial{K} \cap \partial \Omega$) is denoted by $\mathcal{F}$. The following scalar finite element spaces are defined on $\mathcal{T}$ and $\mathcal{F}$:
%
\begin{align}
	W_h   &:= \left\{w_h \in L^2(\mathcal{T}), \hspace{3pt} w_h\lvert_K \hspace{3pt}  \in P_k(K) \hspace{3pt} \forall \hspace{3pt} K \in \mathcal{T} \right\}, \label{eq:wspace_local}\\
	T_h   &:= \left\{\tau_h \in L^2(\mathcal{T}), \hspace{3pt} \tau_h \lvert_K \hspace{3pt}  \in P_l(K) \hspace{3pt} \forall \hspace{3pt} K \in \mathcal{T} \right\}, \label{eq:lagrange_space} \\
	\bar{W}_{h} &:= \left\{\bar{w}_h \in L^{2}(\mathcal{F}), \hspace{3pt} \bar{w}_h \lvert_F \hspace{3pt} \in P_{k} (F) \hspace{3pt} \forall \hspace{3pt} F \in \mathcal{F} \right\},  \label{eq:wspace_global} 
\end{align}
%
in which $P(K)$ and $P(F)$ denote the spaces spanned by Lagrange polynomials on $K$ and $ F $, respectively, and $k\geq 1 $ and $l \geq 0$ indicating the polynomial order. Note that $\fspacel$ in Eq.~\eqref{eq:wspace_local} is equal to Eq.~\eqref{eq:dg_wspace_local}.

Given these function space definitions, the fully-discrete optimality system is obtained after taking variations of the Lagrangian functional Eq.~\eqref{eq:scalar_lagrangian-functional} with respect to $\left(\psi_h, \lambda_h, \bar{\psi}_h  \right) \in \left(W_{h}, T_h  \bar{W}_{h}\right) $ and using a $\theta$-method for the time discretization. The fully-discrete co-state equation in this optimality system reads: 
given the particle field $\psi_p^{n} \in \Psi_t$, the particle positions $\mathbf{x}_p^{n+1} \in \mathcal{X}_t$, and the intermediate field $\xtDiscreteScalar{\psi}{*,n}\in W_h$, find $\left(\xtDiscreteScalar{\psi}{n+1}, \xtDiscreteScalar{\lambda}{n+1}, \xtDiscreteScalar{\bar{\psi}}{n+1} \right) \in  \left( W_h, T_h, \bar{W}_{h} \right)$ such that
%
\begin{subequations} \label{eq:discrete_optimality-adv-diff}
\begin{multline}\label{eq:costate-discrete-adv-diff}
	\sum_{p \in \mathcal{S}_t} 
	\left( \xtDiscreteScalar{\psi}{n+1}(\mathbf{x}_p^{n+1}) - 
	\psi_p^{n} \right) w_h(\mathbf{x}_p^{n+1})   
	-
	\cellsum \lineIntegral{\partial K}{\beta \left( \xtDiscreteScalar{\bar{\psi}}{n+1} - \xtDiscreteScalar{\psi}{n+1} \right) \xDiscreteScalar{w}} 
	+ 
	\sum\limits_{K}^{} \areaIntegral{\partial K}{\zeta \nabla \xtDiscreteScalar{\psi}{n+1} \cdot  \nabla w_h} \\
	+ 
	\areaIntegral{\Omega}{\frac{\xDiscreteScalar{w}}{\Delta t_n} \lambda^{n+1}_h } 
	- 
	\theta \cellsum \areaIntegral{K}{\left(\mathbf{a}^{} \xDiscreteScalar{w} \right) \cdot \nabla \xtDiscreteScalar{\lambda}{n+1} }  
	= 0 
	\hspace{10pt} \forall \; \xDiscreteScalar{w} \in W_h.
\end{multline}
Correspondingly, the fully-discrete counterpart of the state equation becomes:
\begin{multline}\label{eq:state-discrete-adv-diff}
	\areaIntegral{\Omega}{\frac{\xtDiscreteScalar{\psi}{n+1} - \xtDiscreteScalar{\psi}{*,n} }{\Delta t_n} \xDiscreteScalar{\tau}  }    
	- 
	\theta \cellsum \areaIntegral{\cell}{\left(\mathbf{a} \xtDiscreteScalar{\psi}{n+1}\right)  \cdot \nabla \xDiscreteScalar{\tau} }  
	+
	\sum_{K}^{}\lineIntegral{\cellbound}{\mathbf{a} \cdot \unitnormal \: \xtDiscreteScalar{\bar{\psi}}{n+1} \xDiscreteScalar{\tau}} \\
	= 
	(1-\theta) \cellsum \areaIntegral{\cell}{\left(\mathbf{a} \xtDiscreteScalar{\psi}{*,n}\right)  \cdot \GradScalar{\xDiscreteScalar{\tau}} }
	 \hspace{10pt} \forall \; \xDiscreteScalar{\tau} \in T_h.
\end{multline}
Finally, the fully-discrete optimality condition reads
\begin{align}
\sum_{K}^{}\lineIntegral{\partial K}{ \mathbf{a} \cdot \unitnormal \: \xtDiscreteScalar{\lambda}{n+1} \bar{w}_h }  
+
\sum\limits_{K}^{} \lineIntegral{\partial K}{\beta \left( \xtDiscreteScalar{\bar{\psi}}{n+1} - \xtDiscreteScalar{\psi}{n+1} \right) \bar{w}_h }
&= 0 
&& \forall \; \xDiscreteScalar{\bar{w}} \in \bar{W}_{h} \label{eq:optimality-discrete-adv-diff}.
\end{align}
\end{subequations}
%
Note that the Lagrange multiplier $\lambda_h$ and the control variable $\Bar{\psi}_h$ are conveniently chosen at time level $n+1$, which is allowed since these variables are fully-implicit, not requiring differentiation in time. Furthermore, the choice $l=0$ for the polynomial order of the Lagrange multiplier field $\lambda_h$ bears specific advantage in that the terms involving the time-stepping parameter $\theta$ can be dropped, i.e. for $l=0$ the Eq.~\eqref{eq:discrete_optimality-adv-diff} becomes independent of $\theta$. 

Eq.~\eqref{eq:discrete_optimality-adv-diff} can be casted in a $3 \times 3$ block system, see Eq.~\eqref{eq:block_system_cell}. This system of equations is solved for the three unknowns $\left(\xtDiscreteScalar{\psi}{n+1}, \xtDiscreteScalar{\lambda}{n+1}, \xtDiscreteScalar{\bar{\psi}}{n+1} \right) \in  \left( W_h, T_h, \bar{W}_{h} \right)$ via a static condenstaion procedure, see the discussion in Section~\ref{sec:pde_constrained_pm}. 
%




\bibliographystyle{elsarticle-num}
\bibliography{elsarticle-template}

\end{document}